\documentclass[12pt,reqno]{amsart}
\usepackage{graphicx}
\usepackage{amscd}
\usepackage{amsmath}
\usepackage{epsfig}
\usepackage{amsfonts}
\usepackage{amssymb}
\usepackage{hyperref}

\usepackage{pdfpages}
\usepackage[thinc]{esdiff}
\usepackage{accents}
\usepackage{nicefrac}

\providecommand{\U}[1]{\protect\rule{.1in}{.1in}}
\providecommand{\U}[1]{\protect\rule{.1in}{.1in}}
\allowdisplaybreaks
\newtheorem{theorem}{Theorem}
\theoremstyle{plain}

\numberwithin{equation}{section}

\newcommand{\fourvec}[1]{\underaccent{\tilde}{#1}}
\newcommand{\iun}{\mathrm{i}\mkern1mu}

\input{tcilatex}

\begin{document}
\date{\today}
\title[Helicity States of Dirac Free Particles]{Matrix Approach to Helicity
States of Dirac Free Particles\/}
\author{Ben Goren}
\address{School of Mathematical and Statistical Sciences, Arizona State
University, P.~O. Box 871804, Tempe, AZ 85287-1804, U.~S.~A.}
\author{Kamal K. Barley}
\address{Department of Mathematics, Gibbs Hall 307, Allen
University, 1530 Harden St., Columbia, SC 29204, U.~S.~A.}
\author{Sergei K. Suslov$^{1}$}\footnote{Corresponding author. Email: sergei@asu.edu}
\address{School of Mathematical and Statistical Sciences, Arizona State
University, P.~O. Box 871804, Tempe, AZ 85287-1804, U.~S.~A.}


\begin{abstract}
We use elementary matrix algebra to derive the free wave solutions of the Dirac equation
and examine the fundamental concepts of spin, polarization, and helicity states in details.
This consideration can aid readers in studying the mathematical methods of relativistic quantum mechanics.
\end{abstract}

\maketitle

{\scriptsize{Could anything at first sight seem more impractical than a body which is so small that its mass is an insignificant
fraction of the mass of an atom of hydrogen? }}
\begin{flushright}
\scriptsize{{Comment on the electron by J.~J.~Thomson, Nobel Prize in Physics 1906\/} \cite{FKL}}
\end{flushright}
%


%
{\scriptsize{I studied mathematics with passion because I considered it
necessary for the study of physics, \it{to which I want to dedicate myself
exclusively\/}.}}
\begin{flushright}
\scriptsize{{Comment on his mathematical education by Enrico~Fermi, Nobel Prize in Physics 1938\/} \cite{Segre}}
\end{flushright}

\section{Introduction\/}

Although the revolutionary nonrelativistic Schr{\"o}dinger equation allowed us
to explain experimental spectra of atoms and molecules,
the magnificent relativistic Dirac equation laid the sound foundation of the electron's spin,
predicted antimatter, and paved the way to the quantum fields theory
(see, for example, \cite{AkhBer}, \cite{Barleyetal2021}, \cite{BerLifPit}, \cite{Bo:Shi}, \cite{LaLif3}, \cite{Wilczek},
and the references therein).
The electron has been, ever since its discovery at the end of the nineteenth century, subject to intense theoretical and experimental investigations;
say, from radiative corrections of the electron magnetic moment \cite{Aoyamaetal12}, \cite{Hanneketal08}, \cite{FKL}, \cite{Schwinger48}
to testing quantum electrodynamics in strong fields on hydrogen- and helium-like uranium \cite{Gum05}, \cite{Gum07} and measuring hyperfine interval in atomic hydrogen (see recent article \cite{Bullis2023}).
The results of advanced theoretical calculations of quantum electrodynamic effects \cite{Shab02}, \cite{Shabetal06}, \cite{Sturmetal2017}, \cite{Suslov09}
enter as essential input parameters and preliminary estimates in the evaluation of the experimental data \cite{FKL}, \cite{Sturmetal2014}.
Nowadays, physicists gained a deeper understanding of the electron as an elementary particle.
%

From a pedagogical point of view,
a set of $2 \times 2\/$ matrices is simple enough to operate, which forms
the endowment of every introduction to linear algebra.
In nonrelativistic quantum mechanics, the $2 \times 2\/$ matrices and
the corresponding $\mathbb{C}^2$ vectors (spinors) are
primary
to Wolfgang Pauli's theory of spin \cite{LaLif3}, \cite{Merz},
\cite{Schiff}, \cite{Steane}, \cite{Varshalovich1988}.
Solving the Dirac equation for a free particle, a similar ansatz can be used, namely:
$4 \times 4\/$ matrices in a certain $2 \times 2\/$ partitioned block
form; these are almost as easy to work with as $2 \times 2\/$ matrices.
In this article, we shall utilize this simplicity to show that eigenvalues
and corresponding eigenvectors (bi-spinors) of Dirac's free particles,
as well as, their polarization density matrices
can be found in a compact form.
In our opinion, this is essential for a better understanding of the relativistic concept of spin.

In the following section, we introduce the basics of the Dirac
equation (for a free particle). Then, the familiar standard
solutions are verified by matrix multiplication. Next, we correct
a mistake in Fermi's lecture notes \cite{Fermi}
(see {\cite{Bruzz}}, {\cite{Fermi}}, and {\cite{Segre}} regarding the original Fermi's teaching style).
The nonrelativistic and relativistic helicity states of a free Dirac's
particle is discussed in the penultimate section, over again,
with the aid of simple matrix algebra. We conclude by introducing
the corresponding polarization density matrices and also discuss
the property of charge conjugation. Appendix~A presents a summary
of some relevant theorems from the theory of matrices; and
Appendix~B contains our slightly edited typeset version of Fermi's
lecture on the relativistic electron, together with his original
hand-written notes, for the reader's convenience.
Appendix~C holds an original abstract of the talk by Michel and Wightman \cite{MicheWightman55}.

We hope that our somewhat informal presentation may help beginners to enjoy the
study of the mathematics of relativistic quantum mechanics. To this
end, we include important details of calculations that are
usually omitted elsewhere. Computer algebra methods are useful for verification
(our complementary Mathematica notebook is available by request).
This approach is motivated by a course in quantum mechanics which the third-named
author was teaching at Arizona State University for more than two
decades (see also \cite{Barleyetal2021} and \cite{Susetal20}).

\section{Dirac Equation\/}

The relativistic wave equation of Dirac for a free
particle with spin $1/2$, with the electron as the
archetypal example, is given by \cite{AkhBer},
 \cite{BerLifPit}, \cite {BBQED},
\cite{Bj:Dr}, \cite{Bo:Shi}, \cite{Dirac}, \cite{DiracII}, \cite
{DiracQM}, \cite{Fermi}, \cite{Kr:Lan:Sus16}, \cite{Merz},
\cite{Moskalev}, \cite{PeskSchroe}, \cite{Rose61} \cite{Schiff}, \cite{Steane},
\cite{Susetal20}, \cite{Wilczek}:
\begin{equation}
	\iun \hbar \diffp{}{t} \psi(\mathbf{r}, t)
	=
	\widehat{H} \psi(\mathbf{r}, t),
	\label{d1}
\end{equation}
where $\iun$ is the imaginary unit; $\hslash$ is the
reduced Planck constant; $\mathbf{r}$ is a three-component vector
representing position in space; $t$ is time; and $\psi$ is a
complex-valued function.

Before presenting the Hamiltonian, $\widehat{H}$, of this equation
(which represents the energy of the system), we introduce the
matrix elements from which we will build it.
The standard Pauli matrices are given by
\begin{equation}
	{\sigma}_1 =
	\begin{pmatrix}
		0 & 1 \\
		1 & 0
	\end{pmatrix}
	\qquad
	{\sigma}_2 =
	\begin{pmatrix}
		0 & -\iun \\
		\iun & 0
	\end{pmatrix}
	\qquad
	{\sigma}_3 =
	\begin{pmatrix}
		1 & 0 \\
		0 & -1
	\end{pmatrix}.
	\label{d4}
\end{equation}
Note that these correspond with the $x=x_1$, $y=x_2$, and $z=x_3$ axes
(respectively) of three-dimensional Euclidean space.
Also, let
\begin{equation}
	\mathbf{0} =
	\begin{pmatrix}
		0 & 0 \\
		0 & 0
	\end{pmatrix}
	\qquad
	\operatorname{and}
	\qquad
	\mathbf{1} = {I}_2 =
	\begin{pmatrix}
		1 & 0 \\
		0 & 1
	\end{pmatrix} .
	\label{d01}
\end{equation}
From those $2 \times 2$ matrices as building blocks, we construct
several $4 \times 4$ matrices. For each $k$ in the set $\{1, 2, 3\}$, let
\begin{equation}
	{\alpha}_k
	=
	\begin{pmatrix}
		\mathbf{0} & {\sigma}_k \\
		{\sigma}_k & \mathbf{0}
	\end{pmatrix} , \qquad 
{\beta} 	=
	\begin{pmatrix}
		\mathbf{1} & \mathbf{0} \\
		\mathbf{0} & \mathbf{-1}
	\end{pmatrix} .
	\label{d3}
\end{equation}
For example, we may expand (see also equations (34.15) in Appendix~B):
\begin{equation}
	{\alpha_2} =
	\begin{pmatrix}
		0 & 0 & 0 & -\iun \\
		0 & 0 & \iun & 0 \\
		0 & -\iun & 0 & 0 \\
		\iun & 0 & 0 & 0
	\end{pmatrix} , \qquad \beta =
\begin{pmatrix}
		1 & 0 & 0 & 0 \\
		0 & 1 & 0 & 0 \\
		0 & 0 & -1 & 0 \\
		0 & 0 & 0 & -1
	\end{pmatrix} .
\end{equation}
%
%

We may now give the Hamiltonian of the above evolutionary Schr{\"o}dinger-type
equation as follows
\begin{equation}
	\widehat{H}
	=
	c
	\left(
		\boldsymbol{\alpha}
		\cdot
		\widehat{\boldsymbol{p}}
	\right)
	+mc^{2} {\beta},
	\label{d2}
\end{equation}
where $\boldsymbol{\alpha} \cdot \widehat{\boldsymbol{p}} =
{\alpha_1} \widehat{p}_1 + {\alpha_2}
\widehat{p}_2 + {\alpha _{3}} \widehat{p}_{3}$ with the
linear momentum operator $\widehat{\boldsymbol{p}}=-\iun\hbar
\nabla$, and where $p_1$, $p_2$, and $p_3$ are the components of
linear momentum in the $x$, $y$, and $z$ directions. As usual,
$m$ is the rest mass and $c$ is the speed of light in a vacuum.
(Throughout this article we will generally use the same notation
as in \cite{Kr:Lan:Sus16}; see also the references therein.)

The relativistic electron has a four component (bi-spinor)
complex-valued wave function
\begin{equation}
	\psi \left( \mathbf{r},t\right) =
	\begin{pmatrix}
		\psi_{1}\left( \mathbf{r},t\right) \\
		\psi_{2}\left( \mathbf{r},t\right) \\
		\psi_{3}\left( \mathbf{r},t\right) \\
		\psi_{4}\left( \mathbf{r},t\right)
	\end{pmatrix}.
	\label{d4a}
\end{equation}
As a result, the Dirac equation (\ref{d1}) with the Hamiltonian (\ref{d2}) is a matrix equation
that is equivalent to a system of four first order partial
differential equations.

The standard harmonic plane wave solution for a frame of reference
that is at rest has the form:
\begin{equation}
	\psi
	=
	\psi \left(\mathbf{r}, t\right)
	=
	e^{
		\frac{ \iun }{\hbar}
		(\boldsymbol{p} \cdot \boldsymbol{r}-Et)
	}u
	=
	\exp \left(
		\frac{ \iun }{\hbar}
		(\boldsymbol{p\cdot r}-Et)
	\right)
	\begin{pmatrix}
		u_{1} \\
		u_{2} \\
		u_{3} \\
		u_{4}
	\end{pmatrix}
	,
	\label{d7}
\end{equation}
where $u$ is a constant four vector (bi-spinor). We choose
the eigenfunctions with definite energy and linear momentum of the commuting energy and momentum operators:
%
\begin{equation}
	 \iun \hbar \diffp{}{t} \psi = E \psi,
	\qquad
	\boldsymbol{\widehat{p}} \psi = \boldsymbol{p} \psi.
	\label{d7a}
\end{equation}
Substitution into the Dirac equation results in an eigenvalue
problem:
\begin{equation}
	\left(
		c
		\boldsymbol{\alpha} \cdot \boldsymbol{p}
		+mc^{2}
		{\beta}
	\right)
	u
	=
	Eu,
	\label{d8}
\end{equation}
where $E$ is the total energy and $\boldsymbol{p}$ is the linear momentum of
the free electron.

We may write the same equation in block form using our matrices
from above:
\begin{equation}
	\begin{pmatrix}
		(mc^2-E) \mathbf{1} & c \boldsymbol{\sigma} \cdot \boldsymbol{p} \\
		c \boldsymbol{\sigma} \cdot \boldsymbol{p} & -(mc^2+E) \mathbf{1}
	\end{pmatrix}
	u = 0.
	\label{d9}
\end{equation}
Our immediate goal now is to show that this representation is very
convenient for finding eigenvalues and their corresponding eigenvectors
by means of simple matrix algebra.
(In the following an identity matrix is usually assumed when needed.)

\section{Matrix Algebra Plane Wave Solution\/}

We would like to point out that the eigenvalue problem (\ref{d8})--(\ref{d9}%
) can be easily solved with the help of the following matrix identity:%
\begin{eqnarray}
&&\left(
\begin{array}{cc}
\left( mc^{2}-E\right) \mathbf{1} & c\boldsymbol{\sigma \cdot p} \\
c\boldsymbol{\sigma \cdot p} & -\left( mc^{2}+E\right) \mathbf{1}%
\end{array}%
\right) \left(
\begin{array}{cc}
\left( mc^{2}+E\right) \mathbf{1} & c\boldsymbol{\sigma \cdot p} \\
c\boldsymbol{\sigma \cdot p} & -\left( mc^{2}-E\right) \mathbf{1}%
\end{array}%
\right) \label{ma1} \\
&&\qquad \qquad \qquad \qquad \qquad \qquad \quad \qquad =\left(
m^{2}c^{4}+c^{2}\boldsymbol{p}^{2}-E^{2}\right) \left(
\begin{array}{cc}
\mathbf{1} & \boldsymbol{0} \\
\boldsymbol{0} & \boldsymbol{1}%
\end{array}%
\right) . \notag
\end{eqnarray}%
Here, the first $4\times 4$ partitioned matrix, in the left hand side, is
the same as in (\ref{d9}) and the second one obtained from the latter by the
substitution $E\rightarrow -E.$ They do commute.

Our second observation is the following: a `mixed' partitioned matrix, where
the first two column of the second matrix are combined with the last two
column of the first one, has a diagonal square, namely,
\begin{eqnarray}
&&\left(
\begin{array}{cc}
\left( mc^{2}+E\right) \mathbf{1} & c\boldsymbol{\sigma \cdot p} \\
c\boldsymbol{\sigma \cdot p} & -\left( mc^{2}+E\right) \mathbf{1}%
\end{array}%
\right) \left(
\begin{array}{cc}
\left( mc^{2}+E\right) \mathbf{1} & c\boldsymbol{\sigma \cdot p} \\
c\boldsymbol{\sigma \cdot p} & -\left( mc^{2}+E\right) \mathbf{1}%
\end{array}%
\right) \label{ma2} \\
&&\qquad \qquad \qquad \qquad \qquad \qquad \qquad =\left( \left(
mc^{2}+E\right) ^{2}+c^{2}\boldsymbol{p}^{2}\right) \left(
\begin{array}{cc}
\mathbf{1} & \boldsymbol{0} \\
\boldsymbol{0} & \boldsymbol{1}%
\end{array}%
\right) . \notag
\end{eqnarray}%
It is easy to show that these two elementary facts give a standard solution
of the above eigenvalue problem.

Let us verify both identities. Indeed, by the multiplication rule for
partitioned matrices (\ref{A1}), one gets%
\begin{eqnarray}
\!\!\! &&\!\!\!\left(
\begin{array}{cc}
\left( mc^{2}-E\right) \mathbf{1} & c\boldsymbol{\sigma \cdot p} \\
c\boldsymbol{\sigma \cdot p} & -\left( mc^{2}+E\right) \mathbf{1}%
\end{array}%
\right) \left(
\begin{array}{cc}
\left( mc^{2}+E\right) \mathbf{1} & c\boldsymbol{\sigma \cdot p} \\
c\boldsymbol{\sigma \cdot p} & -\left( mc^{2}-E\right) \mathbf{1}%
\end{array}%
\right) \label{ma3} \\
\!\!\!\!\! &=&\!\!\!\!\left(
\begin{array}{cc}
\left( m^{2}c^{4}-E^{2}\right) \mathbf{1+}c^{2}\left( \boldsymbol{\sigma
\cdot p}\right) ^{2} & \!\!\!\left( mc^{2}-E\right) \left( c\boldsymbol{%
\sigma \cdot p}\right) -\left( c\boldsymbol{\sigma \cdot p}\right) \left(
mc^{2}-E\right) \\
\!\!\left( c\boldsymbol{\sigma \cdot p}\right) \left( mc^{2}+E\right)
-\left( mc^{2}+E\right) \left( c\boldsymbol{\sigma \cdot p}\right) &
c^{2}\left( \boldsymbol{\sigma \cdot p}\right) ^{2}+\left(
m^{2}c^{4}-E^{2}\right) \mathbf{1}%
\end{array}%
\right) \!. \notag
\end{eqnarray}%
But $\left( \boldsymbol{\sigma \cdot p}\right) ^{2}=\boldsymbol{p}^{2}%
\mathbf{1}$ and, in view of diagonalization of the product, the first
identity follows. The second one can be verified in a similar fashion.

According to our first identity (\ref{ma1}), all four column vectors of the
second matrix are automatically give some eigenvectors provided%
\begin{equation}
m^{2}c^{4}+c^{2}\boldsymbol{p}^{2}=E^{2},\qquad \text{or}\qquad E=E_{\pm
}=\pm R,\qquad R=\sqrt{c^{2}\boldsymbol{p}^{2}+m^{2}c^{4}}. \label{ma4}
\end{equation}%
But the rank of this matrix is only two due to Theorem~4 on p.~47 in \cite%
{GantmacherMat} (see also Appendix~A for the reader's convenience) and only
two of those column vectors are linearly independent. Indeed, in this case,%
\begin{equation}
\left(
\begin{array}{cc}
\left( mc^{2}+E\right) \mathbf{1} & c\boldsymbol{\sigma \cdot p} \\
c\boldsymbol{\sigma \cdot p} & -\left( mc^{2}-E\right) \mathbf{1}%
\end{array}%
\right) =\left(
\begin{array}{cc}
A & B \\
C & D%
\end{array}%
\right) , \label{ma4a}
\end{equation}%
and the required identity $D=CA^{-1}B$ is satisfied.

But our second normalized unitary matrix, namely,%
\begin{equation}
U=\frac{1}{\sqrt{\left( mc^{2}+R\right) ^{2}+c^{2}\boldsymbol{p}^{2}}}\left(
\begin{array}{cc}
\left( mc^{2}+R\right) \mathbf{1} & c\boldsymbol{\sigma \cdot p} \\
c\boldsymbol{\sigma \cdot p} & -\left( mc^{2}+R\right) \mathbf{1}%
\end{array}%
\right) ,\qquad U=U^{\dag },\quad U^{2}=\mathbf{1} \label{ma5}
\end{equation}%
is, obviously, nonsingular and provide all four linearly independent column
vectors. Thus, in an explicit matrix form,
\begin{eqnarray}
U &=&\left( u^{\left( 1\right) },u^{\left( 2\right) },u^{\left( 3\right)
},u^{\left( 4\right) }\right) =\sqrt{\frac{mc^{2}+R}{2R}}\left(
\begin{array}{cc}
\mathbf{1} & \dfrac{c\boldsymbol{\sigma \cdot p}}{mc^{2}+R} \\
\dfrac{c\boldsymbol{\sigma \cdot p}}{mc^{2}+R} & -\mathbf{1}%
\end{array}%
\right) \notag \\
&=&\sqrt{\frac{mc^{2}+R}{2R}}\left(
\begin{array}{cccc}
1 & 0 & \dfrac{cp_{3}}{mc^{2}+R} & \dfrac{c\left( p_{1}- \iun p_{2}\right) }{%
mc^{2}+R} \\
0 & 1 & \dfrac{c\left( p_{1}+ \iun p_{2}\right) }{mc^{2}+R} & \dfrac{-cp_{3}}{%
mc^{2}+R} \\
\dfrac{cp_{3}}{mc^{2}+R} & \dfrac{c\left( p_{1}- \iun p_{2}\right) }{mc^{2}+R} &
-1 & 0 \\
\dfrac{c\left( p_{1}+ \iun p_{2}\right) }{mc^{2}+R} & \dfrac{-cp_{3}}{mc^{2}+R} &
0 & -1%
\end{array}%
\right) . \label{ma6}
\end{eqnarray}%
As a result, in a traditional bi-spinor form, the normalized eigenvectors
are given by%
\begin{equation}
u^{\left( 1\right) }=\sqrt{\frac{mc^{2}+R}{2R}}\left(
\begin{array}{c}
1 \\
0 \\
\dfrac{cp_{3}}{mc^{2}+R} \\
\dfrac{c\left( p_{1}+ \iun p_{2}\right) }{mc^{2}+R}%
\end{array}%
\right) ,\qquad u^{\left( 2\right) }=\sqrt{\frac{mc^{2}+R}{2R}}\left(
\begin{array}{c}
0 \\
1 \\
\dfrac{c\left( p_{1}- \iun p_{2}\right) }{mc^{2}+R} \\
\dfrac{-cp_{3}}{mc^{2}+R}%
\end{array}%
\right) \label{ma7}
\end{equation}%
for the positive energy eigenvalues $E=E_{+}=R=\sqrt{c^{2}\boldsymbol{p}%
^{2}+m^{2}c^{4}}$ (twice) with the projection of the spin on the third axis $%
\pm 1/2,$ in the frame of reference when the particle is at rest, $%
\boldsymbol{p}=\boldsymbol{0},$ respectively, whereas the normalized
eigenvectors:%
\begin{equation}
u^{\left( 3\right) }=\sqrt{\frac{mc^{2}+R}{2R}}\left(
\begin{array}{c}
\dfrac{cp_{3}}{mc^{2}+R} \\
\dfrac{c\left( p_{1}+ \iun p_{2}\right) }{mc^{2}+R} \\
-1 \\
0%
\end{array}%
\right) ,\qquad u^{\left( 4\right) }=\sqrt{\frac{mc^{2}+R}{2R}}\left(
\begin{array}{c}
\dfrac{c\left( p_{1}- \iun p_{2}\right) }{mc^{2}+R} \\
\dfrac{-cp_{3}}{mc^{2}+R} \\
0 \\
-1%
\end{array}%
\right) \label{ma8}
\end{equation}%
correspond to the negative energy eigenvalues $E=E_{-}=-R=-\sqrt{c^{2}%
\boldsymbol{p}^{2}+m^{2}c^{4}}$ (twice), once again with the projection of
the spin on the third axis $\pm 1/2,$ when $\boldsymbol{p}=\boldsymbol{0},$
respectively. (For interpretations of the negative energy eigenvalues, see,
for example, \cite{Bj:Dr}, \cite{Bo:Shi}, \cite{DiracQM}, \cite{Fermi}, \cite%
{Merz}, \cite{Moskalev}, \cite{Rose61}, \cite{Schiff}. Mathematica verification
of the bi-spinors is given in the complementary notebook.)

\textbf{Note.\/ }In the nonrelativistic limit, when $c\rightarrow \infty ,$
one gets%
\begin{equation}
R=mc^{2}\sqrt{1+\frac{\boldsymbol{p}^{2}}{\left( mc\right) ^{2}}}=mc^{2}+%
\frac{\boldsymbol{p}^{2}}{2m}+\ ...\ \label{ma8a}
\end{equation}%
and, therefore,%
\begin{equation}
U=\left( u^{\left( 1\right) },u^{\left( 2\right) },u^{\left( 3\right)
},u^{\left( 4\right) }\right) \rightarrow \left(
\begin{array}{cc}
\mathbf{1} & \boldsymbol{0} \\
\boldsymbol{0} & -\mathbf{1}%
\end{array}%
\right) ,\quad \qquad \text{as\quad }\dfrac{\left\vert \boldsymbol{p}%
\right\vert }{mc}\ll 1.\qquad \qquad \blacksquare \label{ma8b}
\end{equation}

The relativistic bi-spinor solutions (\ref{ma6}) can be verified by a direct
substitution into (\ref{d8}). Up to a constant, say, for $E=E_{+}=+R,$ one
gets%
\begin{eqnarray}
&&\!{}\left(
\begin{array}{cc}
\left( mc^{2}-R\right) \mathbf{1} & c\boldsymbol{\sigma \cdot p} \\
c\boldsymbol{\sigma \cdot p} & -\left( mc^{2}+R\right) \mathbf{1}%
\end{array}%
\right) \left(
\begin{array}{c}
\mathbf{1} \\
\dfrac{c\boldsymbol{\sigma \cdot p}}{mc^{2}+R}%
\end{array}%
\right) \label{ma7a} \\
&&\,=\left(
\begin{array}{c}
\left( mc^{2}-R\right) \mathbf{1}+\dfrac{c^{2}\left( \boldsymbol{\sigma
\cdot p}\right) ^{2}}{mc^{2}+R}=\boldsymbol{0\medskip } \\
c\boldsymbol{\sigma \cdot p}-\left( mc^{2}+R\right) \dfrac{c\boldsymbol{%
\sigma \cdot p}}{mc^{2}+R}=\boldsymbol{0}%
\end{array}%
\right) , \notag
\end{eqnarray}%
and, in a similar fashion, for $E=E_{-}=-R:$
\begin{eqnarray}
&&\left(
\begin{array}{cc}
\left( mc^{2}+R\right) \mathbf{1} & c\boldsymbol{\sigma \cdot p} \\
c\boldsymbol{\sigma \cdot p} & -\left( mc^{2}-R\right) \mathbf{1}%
\end{array}%
\right) \left(
\begin{array}{c}
\dfrac{c\boldsymbol{\sigma \cdot p}}{mc^{2}+R} \\
-\mathbf{1}%
\end{array}%
\right) \label{ma7b} \\
&&\,=\left(
\begin{array}{c}
c\boldsymbol{\sigma \cdot p}-c\boldsymbol{\sigma \cdot p}=\boldsymbol{0} \\
\dfrac{c^{2}\left( \boldsymbol{\sigma \cdot p}\right) ^{2}}{mc^{2}+R}+\left(
mc^{2}-R\right) \mathbf{1}=\boldsymbol{0}%
\end{array}%
\right) . \notag
\end{eqnarray}
In our classification of the spin states above, we have emphasized that $\pm
1/2$ projections of spin on the third axis correspond to the frame of
reference when the particle is at rest, $\boldsymbol{p}=\boldsymbol{0}.$
Unfortunately, these solutions and their interpretation are not Lorentz
invariant. Indeed, the spin state of a moving particle, generally speaking,
does not coincide with the one in the frame of reference when the particle
is at rest (the so-called relativistic spin rotation; see, for example, \cite%
{Led:Lyu2003}, \cite{Moskalev} and references therein). A relativistic
classification of the spin states will be discussed later in terms of the
so-called helicity operator; see section~5.

\textbf{Note.\/ }It worth noting that the matrix manipulation above allows
us to bypass a traditional evaluation of the $4\times 4$ determinant in (\ref%
{d9}), resulting into the fourth order characteristic polynomial.
Nonetheless, in view of the formulas of Schur \cite{GantmacherMat}, namely, (%
\ref{A2}) and (\ref{A3a})-- (\ref{A3b}) in Appendix~A, one can obtain%
\begin{eqnarray}
&&\det \left(
\begin{array}{cc}
\left( mc^{2}-E\right) \mathbf{1} & c\boldsymbol{\sigma \cdot p} \\
c\boldsymbol{\sigma \cdot p} & -\left( mc^{2}+E\right) \mathbf{1}%
\end{array}%
\right) \label{ma9} \\
&=&\det \left( -\left( mc^{2}-E\right) \left( mc^{2}+E\right) \mathbf{1-}%
c^{2}\left( \boldsymbol{\sigma \cdot p}\right) ^{2}\right) \notag \\
&=&\det \left( \left( E^{2}-c^{2}\boldsymbol{p}^{2}-m^{2}c^{4}\right)
\mathbf{1}\right) \notag \\
&=&\det \left(
\begin{array}{cc}
E^{2}-c^{2}\boldsymbol{p}^{2}-m^{2}c^{4} & 0 \\
0 & E^{2}-c^{2}\boldsymbol{p}^{2}-m^{2}c^{4}%
\end{array}%
\right) \notag \\
&=&\left( E^{2}-c^{2}\boldsymbol{p}^{2}-m^{2}c^{4}\right) ^{2}=0. \notag
\end{eqnarray}%
These details are usually omitted in the traditional approach (see, for
example, \cite{Fermi} or \cite{Schiff}). $\ \ \ \blacksquare $

\section{Comments on Fermi's Lecture Notes\/}

According to our calculations, all four bi-spinors (26)--(27) on p.~34-6 for
free spin $1/2$ particle in \cite{Fermi}, see also Appendix~B, correspond to
the positive energy eigenvalues $E=E_{+}=+R=\sqrt{c^{2}\boldsymbol{p}%
^{2}+m^{2}c^{4}}.$ This fact can be verified by a direct substitution into
equation (24) on p.~34-5. For example, in the case of the third bi-spinor $%
u^{\left( 3\right) }$ given by equation (27) in Fermi's notes, one gets, up
to a constant, that%
\begin{eqnarray}
&&\left(
\begin{array}{cccc}
mc^{2} & 0 & cp_{3} & c\left( p_{1}- \iun p_{2}\right) \\
0 & mc^{2} & c\left( p_{1}+ \iun p_{2}\right) & -cp_{3} \\
cp_{3} & c\left( p_{1}- \iun p_{2}\right) & -mc^{2} & 0 \\
c\left( p_{1}+ \iun p_{2}\right) & -cp_{3} & 0 & -mc^{2}%
\end{array}%
\right) \left(
\begin{array}{c}
\dfrac{cp_{3}}{R-mc^{2}} \\
\dfrac{c\left( p_{1}+ \iun p_{2}\right) }{R-mc^{2}} \\
1 \\
0%
\end{array}%
\right) \notag \\
&&\quad =\left(
\begin{array}{c}
cp_{3}\left( \dfrac{mc^{2}}{R-mc^{2}}+1\right) =+R\dfrac{cp_{3}}{R-mc^{2}}
\\
c\left( p_{1}+ \iun p_{2}\right) \left( \dfrac{mc^{2}}{R-mc^{2}}+1\right) =+R%
\dfrac{c\left( p_{1}+ \iun p_{2}\right) }{R-mc^{2}} \\
\dfrac{c^{2}\boldsymbol{p}^{2}}{R-mc^{2}}-mc^{2}=\dfrac{R^{2}-m^{2}c^{4}}{%
R-mc^{2}}-mc^{2}=+R \\
c\left( p_{1}+ \iun p_{2}\right) \dfrac{cp_{3}}{R-mc^{2}}-cp_{3}\dfrac{c\left(
p_{1}+ \iun p_{2}\right) }{R-mc^{2}}=0%
\end{array}%
\right) =+R\left(
\begin{array}{c}
\dfrac{cp_{3}}{R-mc^{2}} \\
\dfrac{c\left( p_{1}+ \iun p_{2}\right) }{R-mc^{2}} \\
1 \\
0%
\end{array}%
\right) . \label{cf1}
\end{eqnarray}%
Further details are left to the reader. The correct result is presented, for
example, in \cite{Schiff} and verified in the complementary Mathematica file.
As one can show, the original Fermi's bi-spinors are linearly dependent because the corresponding determinant is equal to zero (see our complementary Mathematica notebook).

\begin{figure}[h!]
	\includegraphics[width=0.75\linewidth]{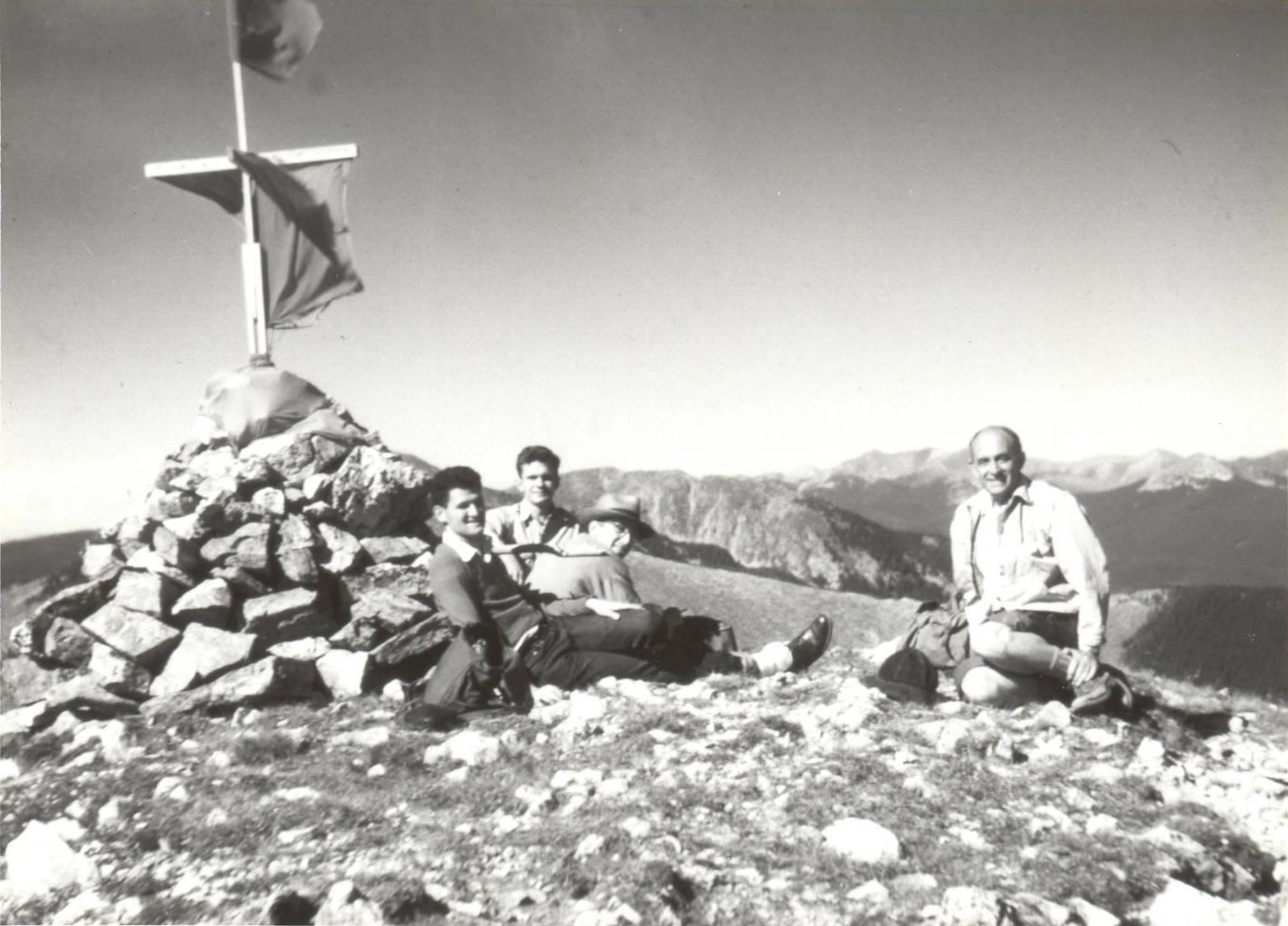}
	\caption{Enrico Fermi hiking with students in New Mexico.
	Courtesy of Peter Lax.}
\end{figure}

\section{Relativistic Helicity States\/}

In the previous sections, de facto, we have used an important property that
the Hamiltonian of a Dirac's free particle (\ref{d2}) commutes with the
linear momentum operator,%
\begin{equation}
\left[ \widehat{H},\ \boldsymbol{\widehat{\boldsymbol{p}}}\right] =0,
\label{s0}
\end{equation}%
in order to define the states with definite values of energy and linear
momentum. But electron has also the spin, or own internal angular momentum
\cite{Mehra1969}. The corresponding integral of motion, related to the
so-called helicity operator, will be discussed here.

\subsection{Helicity Operator}

Introducing standard $4\times 4$ spin matrices and the relativistic spin
operator \cite{Merz}, \cite{Moskalev}, \cite{Schiff}:%
\begin{equation}
\mathbf{\Sigma }=\left(
\begin{array}{cc}
\boldsymbol{\sigma } & \mathbf{0} \\
\mathbf{0} & \boldsymbol{\sigma }%
\end{array}%
\right) ,\qquad \widehat{\boldsymbol{S}}=\frac{1}{2}\boldsymbol{\Sigma },
\label{s1}
\end{equation}%
one can show that%
\begin{equation}
\left[ \widehat{H},\ \boldsymbol{\Sigma }\right] =2 \iun c\left( \boldsymbol{%
\alpha \times \widehat{\boldsymbol{p}}}\right) \neq 0. \label{s2}
\end{equation}%
Indeed, by (\ref{d2}), in components,%
\begin{eqnarray}
\left[ \widehat{H},\ \Sigma _{q}\right] &=&\left[ c\alpha _{r}\ \widehat{%
p_{r}}+mc^{2}\beta ,\ \Sigma _{q}\right] \label{s2ab} \\
&=&c\widehat{p}_{r}\left[ \alpha _{r},\ \Sigma _{q}\right] +mc^{2}\left[
\beta ,\ \Sigma _{q}\right] , \notag
\end{eqnarray}%
where%
\begin{equation}
\left[ \alpha _{r},\ \Sigma _{q}\right] =2 \iun e_{rqs}\alpha _{s},\qquad \left[
\beta ,\ \Sigma _{q}\right] =0, \label{2ba}
\end{equation}%
by the block matrix multiplication (\ref{A1}) and familiar properties of
Pauli's matrices:%
\begin{equation}
\left[ \sigma _{q},\ \sigma _{r}\right] =2 \iun e_{qrs}\sigma _{s}\qquad \left(
q,r,s=1,2,3\right) . \label{s2aba}
\end{equation}%
Here, $e_{qrs}$ is the Levi-Civita symbol \cite{LaLif3}, \cite%
{Varshalovich1988}.

As a result, we obtain
\begin{equation}
\left[ \widehat{H},\ \boldsymbol{\Sigma \cdot \widehat{\boldsymbol{p}}}%
\right] =2 \iun c\left( \boldsymbol{\alpha \times \widehat{\boldsymbol{p}}}%
\right) \boldsymbol{\cdot \widehat{\boldsymbol{p}}=0}, \label{s2a}
\end{equation}%
and the following helicity operator:%
\begin{equation}
\widehat{\Lambda }=\frac{\ \boldsymbol{\widehat{\boldsymbol{S}}\cdot
\widehat{\boldsymbol{p}}}}{\left\vert \boldsymbol{\boldsymbol{p}}\right\vert
}, \label{s4}
\end{equation}%
commutes with the Hamiltonian of Dirac's free particle: $\left[ \widehat{H}%
,\ \widehat{\Lambda }\right] =0.$ Its eigenvalues $\lambda =\pm 1/2$
correspond to the relativistic states with the projections of spin in the
directions $\pm $ $\boldsymbol{\boldsymbol{p}},$ respectively.

For a Dirac particle, only the total angular momentum operator, $\widehat{%
\boldsymbol{J}}=\widehat{\boldsymbol{L}}+\widehat{\boldsymbol{S}},$ $%
\widehat{\boldsymbol{L}}=\mathbf{r}\times \widehat{\boldsymbol{p}},$ namely,
the sum of the orbital angular momentum operator and the spin, does commute
with the Hamiltonian: $\left[ \widehat{H},\ \boldsymbol{J}\right] =0$ (see,
for example, \cite{Moskalev}, \cite{Schiff} for more details). In other
words, neither the orbital angular momentum $\widehat{\boldsymbol{L}},$ nor
spin $\widehat{\boldsymbol{S}},$ are separately integrals of motion; only
the total angular momentum $\widehat{\boldsymbol{J}}=\widehat{\boldsymbol{L}}%
+\widehat{\boldsymbol{S}}$ conserves. Therefore, the plane waves under
consideration, as states with a given energy $E$ and linear momentum $%
\boldsymbol{p},$ cannot have a definite projection of the particle spin $%
\widehat{\boldsymbol{S}}$ on an arbitrary axis, say $z=x_{3}.$ Nonetheless,
as our equations (\ref{s2a})--(\ref{s4}) show, the exceptions are the states
with projection of spin towards and opposite to the direction of the linear
momentum because $\widehat{\boldsymbol{L}}\cdot \widehat{\boldsymbol{p}}%
=\left( \mathbf{r}\times \widehat{\boldsymbol{p}}\right) \cdot \widehat{%
\boldsymbol{p}}=0$ and, therefore, $\widehat{\boldsymbol{J}}\cdot \widehat{%
\boldsymbol{p}}=\widehat{\boldsymbol{S}}\cdot \widehat{\boldsymbol{p}}.$ In
this section, we will construct these Lorentz invariant (covariant) spin
states in details.

\subsection{Nonrelativistic Helicity States}

To get started, one may recall that for the nonrelativistic spin operator
\cite{LaLif3}, \cite{Varshalovich1988}:%
\begin{equation}
\boldsymbol{\widehat{\boldsymbol{s}}}=\frac{1}{2}\boldsymbol{\sigma },
\label{s5}
\end{equation}%
say, in the frame of reference, that is moving together with the particle,
with the spin pointed out in the direction of a unit polarization vector $%
\boldsymbol{\boldsymbol{n}}=\boldsymbol{\boldsymbol{n}}\left( \theta
,\varphi \right) :$%


\begin{eqnarray}
n_{1} &=&\sin \theta \cos \varphi , \notag \\
n_{2} &=&\sin \theta \sin \varphi , \label{s6} \\
n_{3} &=&\cos \varphi , \notag
\end{eqnarray}%
the corresponding spinors satisfy the following eigenvalue problem (helicity
states \cite{Moskalev}, \cite{Varshalovich1988}):%
\begin{equation}
\left( \widehat{\boldsymbol{s}}\boldsymbol{\cdot \boldsymbol{n}}\right) \phi
=\frac{1}{2}\left( \boldsymbol{\boldsymbol{\sigma }\cdot \boldsymbol{n}}%
\right) \phi =\lambda \phi . \label{s7}
\end{equation}%
The standard solutions are%
\begin{equation}
\phi ^{\left( 1/2\right) }\left( \theta ,\varphi \right) =\left(
\begin{array}{c}
\cos \dfrac{\theta }{2}\ e^{- \iun \varphi /2} \\
\sin \dfrac{\theta }{2}\ e^{ \iun \varphi /2}%
\end{array}%
\right) \quad \left( \text{when }\lambda =+\frac{1}{2}\right) ,\qquad \phi
^{\left( 1/2\right) }\left( 0,0\right) =\left(
\begin{array}{c}
1 \\
0%
\end{array}%
\right) ; \label{s8a}
\end{equation}%
and%
\begin{equation}
\phi ^{\left( -1/2\right) }\left( \theta ,\varphi \right) =\left(
\begin{array}{c}
-\sin \dfrac{\theta }{2}\ e^{- \iun \varphi /2} \\
\cos \dfrac{\theta }{2}\ e^{ \iun \varphi /2}%
\end{array}%
\right) \quad \left( \text{when }\lambda =-\frac{1}{2}\right) ,\qquad \phi
^{\left( 1/2\right) }\left( 0,0\right) =\left(
\begin{array}{c}
0 \\
1%
\end{array}%
\right) \label{s8b}
\end{equation}%
with the projection of spin in the direction $\pm \boldsymbol{\boldsymbol{n}}%
\left( \theta ,\varphi \right) ,$ respectively.

As a result, in compact form,%
\begin{equation}
\left( \boldsymbol{\boldsymbol{\sigma }\cdot \boldsymbol{n}}\right) \phi
^{\left( \pm 1/2\right) }=\pm \phi ^{\left( \pm 1/2\right) }, \label{s8bb}
\end{equation}%
and, vice versa,%
\begin{equation}
\left( \phi ^{\left( \pm 1/2\right) }\right) ^{\dag }\left( \boldsymbol{%
\boldsymbol{\sigma }}\phi ^{\left( \pm 1/2\right) }\right) =\pm \boldsymbol{%
\boldsymbol{n}}=\pm \boldsymbol{\boldsymbol{n}}\left( \theta ,\varphi
\right) . \label{s8bc}
\end{equation}%
These facts can be verified by direct substitutions (details are left to the
reader).

Both nonrelativistic helicity states can be thought of as the columns of the
following $2\times 2$ unitary matrix;%
\begin{equation}
\Phi =\left( \phi ^{\left( 1/2\right) },\ \phi ^{\left( -1/2\right) }\right)
=\left(
\begin{array}{cc}
\cos \dfrac{\theta }{2}e^{- \iun \varphi /2} & -\sin \dfrac{\theta }{2}%
e^{- \iun \varphi /2} \\
\sin \dfrac{\theta }{2}e^{ \iun \varphi /2} & \cos \dfrac{\theta
	}{2}e^{ \iun \varphi
/2}%
\end{array}%
\right) ,\qquad \Phi \ \Phi ^{\dag }=\boldsymbol{1}, \label{s8c}
\end{equation}%
which is related to the spin $1/2$ finite rotation matrix \cite{LaLif3},
\cite{Varshalovich1988}. Introducing also,%
\begin{equation}
\widetilde{\Phi }=\left( \phi ^{\left( 1/2\right) },\ -\phi ^{\left(
-1/2\right) }\right) =\left(
\begin{array}{cc}
\cos \dfrac{\theta }{2}e^{- \iun \varphi /2} & \sin \dfrac{\theta }{2}%
e^{- \iun \varphi /2} \\
\sin \dfrac{\theta }{2}e^{ \iun \varphi /2} & -\cos \dfrac{\theta
	}{2}e^{ \iun \varphi
/2}%
\end{array}%
\right) ,\qquad \widetilde{\Phi }\ \widetilde{\Phi }^{\dag }=\boldsymbol{1},
\label{s8d}
\end{equation}%
we will establish the following factorization properties of the helicity
matrix:%
\begin{equation}
\boldsymbol{\boldsymbol{\sigma }\cdot \boldsymbol{n}}=\widetilde{\Phi }\
\Phi ^{\dag }=\Phi \ \widetilde{\Phi }\ ^{\dag }. \label{s8e}
\end{equation}%
Indeed, rewriting in the right-hand side, say $\widetilde{\Phi }\ \Phi
^{\dag },$ in explicit form, one gets:%
\begin{eqnarray}
&&\left(
\begin{array}{cc}
\cos \dfrac{\theta }{2}e^{- \iun \varphi /2} & \sin \dfrac{\theta }{2}%
e^{- \iun \varphi /2} \\
\sin \dfrac{\theta }{2}e^{ \iun \varphi /2} & -\cos \dfrac{\theta
	}{2}e^{ \iun \varphi
/2}%
\end{array}%
\right) \left(
\begin{array}{cc}
\cos \dfrac{\theta }{2}e^{ \iun \varphi /2} & \sin \dfrac{\theta
	}{2}e^{- \iun \varphi
/2} \\
-\sin \dfrac{\theta }{2}e^{ \iun \varphi /2} & \cos \dfrac{\theta }{2}%
e^{- \iun \varphi /2}%
\end{array}%
\right) \label{s8f} \\
&&\qquad \qquad \qquad \qquad \qquad =\left(
\begin{array}{cc}
\cos \theta & \sin \theta \ e^{- \iun \varphi } \\
\sin \theta \ e^{ \iun \varphi } & -\cos \theta%
\end{array}%
\right) =\boldsymbol{\boldsymbol{\sigma }\cdot \boldsymbol{n}}, \notag
\end{eqnarray}%
by the familiar double-angle trigonometric identities. As a result, the
eigenvalue problem (\ref{s7}), or (\ref{s8bb}), is verified:%
\begin{equation}
\left( \boldsymbol{\boldsymbol{\sigma }\cdot \boldsymbol{n}}\right) \Phi
=\left( \widetilde{\Phi }\ \Phi ^{\dag }\right) \Phi =\widetilde{\Phi }%
\left( \Phi ^{\dag }\Phi \right) =\widetilde{\Phi }, \label{s8g}
\end{equation}%
by a simple matrix multiplication. Thus,%
\begin{equation}
\left( \boldsymbol{\boldsymbol{\sigma }\cdot \boldsymbol{n}}\right) \Phi =%
\widetilde{\Phi },\qquad \left( \boldsymbol{\boldsymbol{\sigma }\cdot
\boldsymbol{n}}\right) \widetilde{\Phi }=\Phi , \label{s8h}
\end{equation}%
say, due to a familiar relation $\left( \boldsymbol{\sigma \cdot n}\right)
^{2}=\mathbf{1}.$ (We haven't been able to find these calculations in the
available literature.)

\textbf{Note}. Introducing the spin one-half polarization density matrix as
follows%
\begin{equation}
\rho _{rs}^{\left( \lambda \right) }\left( \boldsymbol{\boldsymbol{n}}%
\right) :=\phi _{r}^{\left( \lambda \right) }\left( \boldsymbol{\boldsymbol{n%
}}\right) \left[ \phi _{s}^{\left( \lambda \right) }\left( \boldsymbol{%
\boldsymbol{n}}\right) \right] ^{\ast }\qquad \left( r,s=1,2\right) ,
\label{s8aa}
\end{equation}
one gets%
\begin{equation}
\rho ^{\left( \pm 1/2\right) }\left( \boldsymbol{\boldsymbol{n}}\right) =%
\frac{1}{2}\left( \mathbf{1}\pm \boldsymbol{\sigma \cdot n}\right) .
\label{s8ab}
\end{equation}%
(More details can be found in \cite{LaLif3}, \cite{Moskalev}, \cite%
{Varshalovich1988}.) \ \ \ $\blacksquare $

\subsection{Relativistic Plane Waves}

Let us extend the original solutions (\ref{ma6}) and analyze them from a
different perspective. For a free relativistic particle with the linear
momentum $\boldsymbol{\boldsymbol{p}}$ and energy $E,$ one can look for
solutions of (\ref{d9}) in the block form%
\begin{equation}
v\left( p\right) =C\left(
\begin{array}{c}
\phi \\
\chi%
\end{array}%
\right) , \label{s9}
\end{equation}%
where $C$ is a constant and $\phi ,$ $\chi $ are two-component spinors:%
\begin{equation}
\phi =\left(
\begin{array}{c}
\phi _{1} \\
\phi _{2}%
\end{array}%
\right) ,\qquad \chi =\left(
\begin{array}{c}
\chi _{1} \\
\chi _{2}%
\end{array}%
\right) . \label{s10}
\end{equation}%
Then%
\begin{eqnarray}
\left( mc^{2}-E\right) \phi +c\left( \boldsymbol{\sigma \cdot p}\right) \chi
&=&0, \label{s11} \\
c\left( \boldsymbol{\sigma \cdot p}\right) \phi -\left( mc^{2}+E\right) \chi
&=&0. \notag
\end{eqnarray}%
From the second (first) equation:%
\begin{equation}
\chi =\frac{c\left( \boldsymbol{\sigma \cdot p}\right) }{mc^{2}+E}\phi
\qquad \left( \phi =\frac{c\left( \boldsymbol{\sigma \cdot p}\right) }{%
E-mc^{2}}\chi \right) \label{s12}
\end{equation}%
and the substitution of this relation into the first (second) equation gives
the relativistic spectrum (\ref{ma4}) for a spinor $\phi $ (correspondingly,
$\chi $) related to an arbitrary polarization vector $\boldsymbol{%
\boldsymbol{n}},$ in the frame of reference where the particle is at rest.
[Substitution of the second relation (\ref{s12}) into the first one results
in an identity on the spectrum (\ref{ma4}), and vise versa.]

As a result, the general solutions under consideration have the forms \cite%
{Moskalev}:%
\begin{equation}
v\left( p\right) =C\left(
\begin{array}{c}
\phi \\
\dfrac{c\left( \boldsymbol{\sigma \cdot p}\right) }{mc^{2}+E}\phi%
\end{array}%
\right) \quad \qquad \left( \text{and }D\left(
\begin{array}{c}
\dfrac{c\left( \boldsymbol{\sigma \cdot p}\right) }{mc^{2}-E}\chi \\
-\chi%
\end{array}%
\right) \right) . \label{s13}
\end{equation}

\textbf{Note}. These formulas can also be obtained by the (inverse) Lorentz
boost towards to the direction $\boldsymbol{\boldsymbol{l}}=\boldsymbol{%
\boldsymbol{p/}}\left\vert \boldsymbol{\boldsymbol{p}}\right\vert $ from the
frame of reference when the particle is at rest. Indeed, the system (\ref%
{s11}) has a trivial solution $E=mc^{2},$ $\boldsymbol{p}=\boldsymbol{0}$
and $\chi =0$ with an arbitrary spinor $\phi .$ The above transformation of
energy and momentum takes the form:%
\begin{equation}
\left(
\begin{array}{c}
E/c \\
\left\vert \boldsymbol{\boldsymbol{p}}\right\vert%
\end{array}%
\right) =\left(
\begin{array}{cc}
\cosh \vartheta & \sinh \vartheta \\
\sinh \vartheta & \cosh \vartheta%
\end{array}%
\right) \left(
\begin{array}{c}
mc \\
0%
\end{array}%
\right) , \label{s13aa}
\end{equation}%
or $E=mc^{2}\cosh \vartheta $ and $\left\vert \boldsymbol{\boldsymbol{p}}%
\right\vert =mc\sinh \vartheta .$ For positive energy eigenvalues, one gets
\begin{eqnarray}
v\left( p\right) &=&S_{L^{-1}}v\left( \boldsymbol{\boldsymbol{p}}=\mathbf{0}%
\right) =e^{-\vartheta \left( \boldsymbol{\sigma \cdot l}\right) /2}v\left(
0\right) =\left( \cosh \frac{\vartheta }{2}+\left( \boldsymbol{\alpha \cdot l%
}\right) \sinh \frac{\vartheta }{2}\right) v\left( 0\right) \label{s13a} \\
&=&\left(
\begin{array}{cc}
\cosh \dfrac{\vartheta }{2} & \left( \boldsymbol{\sigma \cdot l}\right)
\sinh \dfrac{\vartheta }{2} \\
\left( \boldsymbol{\sigma \cdot l}\right) \sinh \dfrac{\vartheta }{2} &
\cosh \dfrac{\vartheta }{2}%
\end{array}%
\right) \left(
\begin{array}{c}
\phi \\
0%
\end{array}%
\right) =\cosh \dfrac{\vartheta }{2}\left(
\begin{array}{c}
\phi \\
\left( \boldsymbol{\sigma \cdot l}\right) \tanh \dfrac{\vartheta }{2}\ \phi%
\end{array}%
\right) \notag
\end{eqnarray}%
(see, for example, \cite{BerLifPit}, \cite{Kr:Lan:Sus16}, \cite{Kr:Lan:Sus60}%
, \cite{Moskalev}, \cite{PeskSchroe}, \cite{Steane2012}, \cite{Steane} for
more details). Here,%
\begin{equation}
\cosh \dfrac{\vartheta }{2}=\sqrt{\frac{\cosh \vartheta +1}{2}}=\sqrt{\frac{%
E+mc^{2}}{2mc^{2}}},\qquad \tanh \dfrac{\vartheta }{2}=\frac{\sinh \vartheta
}{\cosh \vartheta +1}=\frac{c\left\vert \boldsymbol{\boldsymbol{p}}%
\right\vert }{E+mc^{2}}. \label{s13ab}
\end{equation}%
Once again, we obtain the first equation (\ref{s13}) with $C=\sqrt{\left(
mc^{2}+E\right) /2mc^{2}}.$ \ \ $\ \blacksquare $

\subsection{Relativistic Polarization Vector}

We follow \cite{AkhBer}, \cite{Moskalev} with somewhat different details. In
covariant form, the Dirac equation (\ref{d1})-(\ref{d2}) can be written as
follows%
\begin{equation}
\left( \gamma ^{\mu }\widehat{p}_{\mu }-mc\right) \psi =0,\qquad \widehat{p}%
_{\mu }= \iun \hbar \frac{\partial }{\partial x^{\mu }}, \label{rd1}
\end{equation}%
where $x^{\mu }=\left( ct,\boldsymbol{r}\right) ,$ $\gamma ^{\mu }=\left(
\gamma ^{0},\boldsymbol{\gamma }\right) ,$ and%
\begin{equation}
\gamma ^{0}=\beta =\left(
\begin{array}{cc}
\mathbf{1} & \mathbf{0} \\
\mathbf{0} & -\mathbf{1}%
\end{array}%
\right) ,\qquad \boldsymbol{\gamma }=\beta \boldsymbol{\alpha }=\left(
\begin{array}{cc}
\mathbf{0} & \mathbf{\sigma } \\
-\mathbf{\sigma } & \mathbf{0}%
\end{array}%
\right) . \label{rd2}
\end{equation}%
Throughout the article, we use Einstein's summation convention:%
\begin{equation}
a^{\mu }b_{\mu }=g_{\mu \nu }a^{\mu }b^{\nu
}=a^{0}b_{0}+a^{1}b_{1}+a^{2}b_{2}+a^{3}b_{3}=a^{0}b_{0}-\boldsymbol{a\cdot b%
}, \label{rd3}
\end{equation}%
when $a^{\mu }=\left( a^{0},\boldsymbol{a}\right) $ and $b_{\mu }=g_{\mu \nu
}b^{\nu }=\left( b^{0},-\boldsymbol{b}\right) ,$ unless stated otherwise
(see, for example, \cite{Kr:Lan:Sus16} and \cite{Moskalev} for more
details). Here, $g_{\mu \nu }=g^{\mu \nu }=$diag$(1,-1,-1,-1)$ is the
pseudo-Euclidean metric of Minkowski space and the familiar anticommutation
relation holds%
\begin{equation}
\gamma ^{\mu }\gamma ^{\nu }+\gamma ^{\nu }\gamma ^{\mu }=2g^{\mu \nu
}\qquad \left( \mu ,\nu =0,1,2,3\right) . \label{rd3a}
\end{equation}

The bi-spinor wave function for a free Dirac particle, corresponding to a
given energy-momentum four vector, $p_{\mu }=\left( E/c,-\boldsymbol{p}%
\right) $ with the relativistic invariant $p_{\mu }p^{\mu }=\left(
E/c\right) ^{2}-\boldsymbol{p}^{2}=m^{2}c^{2}$ in covariant form, is given by%
\begin{equation}
\psi \left( x\right) =u\left( p\right) e^{- \iun \left( p_{\mu }x^{\mu }\right)
/\hbar },\qquad E=+R=\sqrt{c^{2}\boldsymbol{p}^{2}+m^{2}c^{4}}>0,
\label{rd4}
\end{equation}%
and the standard normalization is as follows%
\begin{equation}
\overline{u}\left( p\right) u\left( p\right) =1,\qquad \overline{u}\left(
p\right) =u^{\dag }\left( p\right) \gamma ^{0}. \label{rd5}
\end{equation}%
The matrix equations for $u\left( p\right) $ and $\overline{u}\left(
p\right) $ take the form \cite{Moskalev}:%
\begin{equation}
\left( \gamma ^{\mu }p_{\mu }-mc\right) u\left( p\right) =0,\qquad \overline{%
u}\left( p\right) \left( \gamma ^{\mu }p_{\mu }-mc\right) =0. \label{rd5a}
\end{equation}%
The normalized bi-spinor is given by%
\begin{equation}
u\left( p\right) =\sqrt{\frac{E+mc^{2}}{2mc^{2}}}\left(
\begin{array}{c}
\phi \\
\dfrac{c\left( \boldsymbol{\sigma \cdot p}\right) }{mc^{2}+E}\phi%
\end{array}%
\right) , \label{rd6}
\end{equation}%
for an arbitrary normalized spinor $\phi :$%
\begin{equation}
\phi ^{\dag }\phi =\left( \phi _{1}^{\ast },\phi _{2}^{\ast }\right) \left(
\begin{array}{c}
\phi _{1} \\
\phi _{2}%
\end{array}%
\right) =\left\vert \phi _{1}\right\vert ^{2}+\left\vert \phi
_{2}\right\vert ^{2}=1. \label{rd7}
\end{equation}%
Indeed, by definition (\ref{rd5}),%
\begin{eqnarray}
&&\left( \phi ^{\dag },\phi ^{\dag }\dfrac{c\left( \boldsymbol{\sigma \cdot p%
}\right) }{mc^{2}+E}\right) \left(
\begin{array}{cc}
\mathbf{1} & \mathbf{0} \\
\mathbf{0} & -\mathbf{1}%
\end{array}%
\right) \left(
\begin{array}{c}
\phi \\
\dfrac{c\left( \boldsymbol{\sigma \cdot p}\right) }{mc^{2}+E}\phi%
\end{array}%
\right) \label{rd8} \\
&&\,=\left( \phi ^{\dag },-\phi ^{\dag }\dfrac{c\left( \boldsymbol{\sigma
\cdot p}\right) }{mc^{2}+E}\right) \left(
\begin{array}{c}
\phi \\
\dfrac{c\left( \boldsymbol{\sigma \cdot p}\right) }{mc^{2}+E}\phi%
\end{array}%
\right) =\phi ^{\dag }\phi -\phi ^{\dag }\dfrac{c^{2}\left( \boldsymbol{%
\sigma \cdot p}\right) ^{2}}{\left( mc^{2}+E\right) ^{2}}\phi \notag \\
&&\,\,=\phi ^{\dag }\phi \left( 1-\dfrac{c^{2}\boldsymbol{p}^{2}}{\left(
mc^{2}+E\right) ^{2}}\right) =\frac{2mc^{2}}{E+mc^{2}}=\frac{1}{\left\vert
C\right\vert ^{2}}. \notag
\end{eqnarray}

Let the spin of the particle be in a certain direction $\boldsymbol{n},$ in
the frame when the particle is at rest, namely,%
\begin{equation}
\left( \boldsymbol{\sigma \cdot n}\right) \phi =\phi ,\qquad \phi ^{\dag }%
\boldsymbol{\sigma }\phi =\boldsymbol{n}. \label{rd9}
\end{equation}%
[The last equation gives the expectation values of nonrelativistic spin
operator (\ref{s5}) in this frame of reference.] Introducing the
relativistic polarization (pseudo) vector as follows \cite{AkhBer}, \cite%
{Moskalev}, \cite{Rose61}, \cite{Tolhoek56}:\footnote{%
Note that $\boldsymbol{\alpha }\gamma ^{5}=\mathbf{\Sigma }.$}
\begin{equation}
a^{\mu }=\overline{\psi }\left( x\right) \left( \gamma _{5}\gamma ^{\mu
}\right) \psi \left( x\right) =\overline{u}\left( p\right) \left( \gamma
_{5}\gamma ^{\mu }\right) u\left( p\right) ,\quad \gamma ^{5}=
	\iun \gamma
^{0}\gamma ^{1}\gamma ^{2}\gamma ^{3}=\left(
\begin{array}{cc}
\mathbf{0} & \mathbf{1} \\
\mathbf{1} & \mathbf{0}%
\end{array}%
\right) =-\gamma _{5}, \label{rd9rd}
\end{equation}%
one can derive the components of this polarization four vector $a^{\mu
}=\left( a^{0},\boldsymbol{a}\right) :$%
\begin{equation}
a^{0}=\frac{\left( \boldsymbol{p\cdot n}\right) }{mc},\qquad \boldsymbol{a}=%
\boldsymbol{n}+\frac{\boldsymbol{p}\left( \boldsymbol{p\cdot n}\right) }{%
m\left( E+mc^{2}\right) }. \label{rd10}
\end{equation}%
Here, the following properties of Pauli's matrices,%
\begin{equation}
\left( \boldsymbol{\sigma \cdot p}\right) \left( \boldsymbol{\sigma \cdot n}%
\right) = \iun \left( \boldsymbol{p\times n}\right) \cdot \boldsymbol{\sigma }%
+\left( \boldsymbol{p\cdot n}\right) \label{rd10a}
\end{equation}%
and%
\begin{equation}
\left( \boldsymbol{\sigma \cdot p}\right) \boldsymbol{\sigma }\left(
\boldsymbol{\sigma \cdot p}\right) =2\boldsymbol{p}\left( \boldsymbol{\sigma
\cdot p}\right) -\boldsymbol{p}^{2}\boldsymbol{\sigma }, \label{rd11}
\end{equation}%
should be utilized along with (\ref{rd9}). [These identities can be verified
with the aid of commutator relations (\ref{s2aba}).]

By definition (\ref{rd9rd}), in components,%
\begin{eqnarray}
a^{0} &=&\left\vert C\right\vert ^{2}\left( \phi ^{\dag },-\phi ^{\dag }%
\dfrac{c\left( \boldsymbol{\sigma \cdot p}\right) }{mc^{2}+E}\right) \left(
\begin{array}{cc}
\mathbf{0} & \mathbf{1} \\
-\mathbf{1} & \mathbf{0}%
\end{array}%
\right) \left(
\begin{array}{c}
\phi \\
\dfrac{c\left( \boldsymbol{\sigma \cdot p}\right) }{mc^{2}+E}\phi%
\end{array}%
\right) \label{rd11a} \\
&=&\left\vert C\right\vert ^{2}\left( \phi ^{\dag },-\phi ^{\dag }\dfrac{%
c\left( \boldsymbol{\sigma \cdot p}\right) }{mc^{2}+E}\right) \left(
\begin{array}{c}
\dfrac{c\left( \boldsymbol{\sigma \cdot p}\right) }{mc^{2}+E}\phi \\
-\phi%
\end{array}%
\right) =\frac{2c\left\vert C\right\vert ^{2}}{E+mc^{2}}\phi ^{\dag }\left(
\boldsymbol{\sigma \cdot p}\right) \left( \boldsymbol{\sigma \cdot n}\right)
\phi \notag \\
&=&\dfrac{1}{mc}\phi ^{\dag }\left(  \iun \left( \boldsymbol{p\times n}\right)
\cdot \boldsymbol{\sigma }+\left( \boldsymbol{p\cdot n}\right) \right) \phi =%
\dfrac{1}{mc}\left( \left( \boldsymbol{p\cdot n}\right) + \iun \left( \boldsymbol{%
p\times n}\right) \cdot \boldsymbol{n}\right) =\dfrac{\boldsymbol{p\cdot n}}{%
mc}. \notag
\end{eqnarray}%
In a similar fashion,%
\begin{eqnarray}
\boldsymbol{a} &=&\left\vert C\right\vert ^{2}\left( \phi ^{\dag },-\phi
^{\dag }\dfrac{c\left( \boldsymbol{\sigma \cdot p}\right) }{mc^{2}+E}\right)
\left(
\begin{array}{cc}
\mathbf{\sigma } & \mathbf{0} \\
\mathbf{0} & -\mathbf{\sigma }%
\end{array}%
\right) \left(
\begin{array}{c}
\phi \\
\dfrac{c\left( \boldsymbol{\sigma \cdot p}\right) }{mc^{2}+E}\phi%
\end{array}%
\right) \label{rd11b} \\
&=&\left\vert C\right\vert ^{2}\left( \phi ^{\dag },-\phi ^{\dag }\dfrac{%
c\left( \boldsymbol{\sigma \cdot p}\right) }{mc^{2}+E}\right) \left(
\begin{array}{c}
\mathbf{\sigma }\phi \\
-\mathbf{\sigma }\dfrac{c\left( \boldsymbol{\sigma \cdot p}\right) }{mc^{2}+E%
}\phi%
\end{array}%
\right) \notag \\
&=&\left\vert C\right\vert ^{2}\left( \phi ^{\dag }\mathbf{\sigma }\phi +%
\dfrac{c^{2}}{\left( mc^{2}+E\right) ^{2}}\phi ^{\dag }\left( \boldsymbol{%
\sigma \cdot p}\right) \boldsymbol{\sigma }\left( \boldsymbol{\sigma \cdot p}%
\right) \phi \right) , \notag
\end{eqnarray}%
where%
\begin{equation*}
\phi ^{\dag }\left( \boldsymbol{\sigma \cdot p}\right) \boldsymbol{\sigma }%
\left( \boldsymbol{\sigma \cdot p}\right) \phi =2\boldsymbol{p}\left( \phi
^{\dag }\boldsymbol{\sigma }\phi \right) \boldsymbol{\cdot p}-\boldsymbol{p}%
^{2}\phi ^{\dag }\boldsymbol{\sigma }\phi ,
\end{equation*}%
in view of identity (\ref{rd11}). As a result,%
\begin{eqnarray}
\boldsymbol{a} &=&\left\vert C\right\vert ^{2}\left( \boldsymbol{n}+\dfrac{%
c^{2}}{\left( mc^{2}+E\right) ^{2}}\left( 2\boldsymbol{p}\left( \boldsymbol{%
n\cdot p}\right) -\boldsymbol{p}^{2}\boldsymbol{n}\right) \right)
\label{rd11c} \\
&=&\left\vert C\right\vert ^{2}\left( 1-\frac{c^{2}\boldsymbol{p}^{2}}{%
\left( mc^{2}+E\right) ^{2}}\right) \boldsymbol{n}+\dfrac{2c^{2}\left\vert
C\right\vert ^{2}}{\left( mc^{2}+E\right) ^{2}}\boldsymbol{p}\left(
\boldsymbol{n\cdot p}\right) \notag \\
&=&\boldsymbol{n}+\frac{\boldsymbol{p}\left( \boldsymbol{p\cdot n}\right) }{%
m\left( E+mc^{2}\right) } \notag
\end{eqnarray}%
by (\ref{rd9}), as stated above.

The four vector (\ref{rd9rd})--(\ref{rd10}) has an important feature,
namely, in the frame of reference, when the particle is at rest, $%
\boldsymbol{p}=\boldsymbol{0},$ one gets%
\begin{equation}
a^{0}=0,\qquad \boldsymbol{a}=\boldsymbol{n}, \label{rd12}
\end{equation}%
thus extending the particle nonrelativistic $3D$ polarization (preudo)
vector. And vice versa, the polarization four vector (\ref{rd10}) can be
obtained from the latter one by the Lorentz transformation \cite{BerLifPit}.

Moreover, the four vector $a^{\mu }$ has the following properties:%
\begin{equation}
p_{\mu }a^{\mu }=0,\qquad a^{2}=a_{\mu }a^{\mu }=-\boldsymbol{n}^{2}=-1.
\label{rd12a}
\end{equation}%
Indeed, by definition (\ref{rd9rd}) and Dirac's equations for bi-spinors $%
u\left( p\right) $, $\overline{u}\left( p\right) ,$ namely (\ref{rd5a}), one
gets%
\begin{eqnarray}
p_{\mu }a^{\mu } &=&\overline{u}\left( p\right) \gamma _{5}\left( p_{\mu
}\gamma ^{\mu }\right) u\left( p\right) =\frac{1}{2}\overline{u}\left(
p\right) \left( \gamma _{5}\left( p_{\mu }\gamma ^{\mu }\right) -\left(
p_{\mu }\gamma ^{\mu }\right) \gamma _{5}\right) u\left( p\right) \notag \\
&=&\frac{1}{2}\overline{u}\left( p\right) \left( \gamma _{5}mc-mc\gamma
_{5}\right) u\left( p\right) =0 \label{rd12aa}
\end{eqnarray}%
in view of a familiar anticommutator relation $\gamma ^{\mu }\gamma
_{5}+\gamma _{5}\gamma ^{\mu }=0$ $\left( \mu =0,1,2,3\right) $. [This
covariant orthogonality relation can be directly verified with the help of (%
\ref{rd10}) and/or evaluated in the frame of reference, when the particle is
at rest, see (\ref{rd12}).]

It worth noting that the bi-spinor $u\left( p\right) ,$ corresponding to the
Dirac free particle with the relativistic polarization vector $a^{\mu },$ in
addition to the Dirac equation,%
\begin{equation}
\left( \gamma ^{\mu }p_{\mu }-mc\right) u\left( p\right) =0, \label{rd12b}
\end{equation}%
also does satisfy the following matrix equation,%
\begin{equation}
\left( \gamma _{5}\gamma ^{\mu }a_{\mu }+I\right) u\left( p\right) =0,
\label{rd12c}
\end{equation}%
which can be verified by direct substitution of $u\left( p\right) $ from (%
\ref{rd6}). [The latter can be thought of as a relativistic generalization
of the first equation (\ref{rd9}), because they coincide when $\boldsymbol{p}%
=\boldsymbol{0}.]$ Indeed, up to a given normalization,%
\begin{eqnarray}
&&\left( \gamma _{5}\gamma ^{\mu }a_{\mu }+I\right) u\left( p\right) =\left(
\begin{array}{cc}
\mathbf{1}-\boldsymbol{\sigma \cdot a} & a^{0} \\
-a^{0} & \mathbf{1}+\boldsymbol{\sigma \cdot a}%
\end{array}%
\right) C\left(
\begin{array}{c}
\phi \\
\dfrac{c\left( \boldsymbol{\sigma \cdot p}\right) }{mc^{2}+E}\phi%
\end{array}%
\right) \label{rd12dd} \\
&&\,=C\left(
\begin{array}{c}
\phi -\left( \boldsymbol{\sigma \cdot a}\right) \phi +a^{0}\dfrac{c\left(
\boldsymbol{\sigma \cdot p}\right) }{mc^{2}+E}\phi \\
-a^{0}\phi +\dfrac{c\left( \boldsymbol{\sigma \cdot p}\right) }{mc^{2}+E}%
\phi +\dfrac{c\left( \boldsymbol{\sigma \cdot a}\right) \left( \boldsymbol{%
\sigma \cdot p}\right) }{mc^{2}+E}\phi%
\end{array}%
\right) \notag \\
&&\,=C\left(
\begin{array}{c}
\phi -\left( \boldsymbol{\sigma \cdot n}\right) \phi -\dfrac{\left(
\boldsymbol{\sigma \cdot p}\right) \left( \boldsymbol{p\cdot n}\right) }{%
m\left( mc^{2}+E\right) }\phi +\dfrac{\left( \boldsymbol{p\cdot n}\right)
\left( \boldsymbol{\sigma \cdot p}\right) }{m\left( mc^{2}+E\right) }\phi =0
\\
-\dfrac{\left( \boldsymbol{p\cdot n}\right) }{mc}\phi +c\dfrac{\left(
\boldsymbol{\sigma \cdot p}\right) \left( \boldsymbol{\sigma \cdot n}\right)
+\left( \boldsymbol{\sigma \cdot n}\right) \left( \boldsymbol{\sigma \cdot p}%
\right) }{mc^{2}+E}\phi +\dfrac{\left( \boldsymbol{p\cdot n}\right) c^{2}%
\boldsymbol{p}^{2}}{mc\left( mc^{2}+E\right) ^{2}}\phi \\
=-\dfrac{\left( \boldsymbol{p\cdot n}\right) }{mc}\phi +\dfrac{2c\left(
\boldsymbol{p\cdot n}\right) }{E+mc^{2}}\phi +\dfrac{\left( \boldsymbol{%
p\cdot n}\right) \left( E^{2}-m^{2}c^{4}\right) }{mc\left( E+mc^{2}\right)
^{2}}\phi \\
=\dfrac{\left( \boldsymbol{p\cdot n}\right) }{mc}\left( -1+\dfrac{2mc^{2}}{%
E+mc^{2}}+\dfrac{E-mc^{2}}{E+mc^{2}}\right) \phi =0%
\end{array}%
\right) \notag
\end{eqnarray}%
by (\ref{rd9}) and (\ref{rd10})--(\ref{rd10a}).

It worth mentioning, in conclusion, that the current density four vector for
a free Dirac particle with a definite energy and linear momentum is given by%
\begin{equation}
j^{\mu }=\overline{\psi }\left( x\right) \left( \gamma ^{\mu }\right) \psi
\left( x\right) =\overline{u}\left( p\right) \left( \gamma ^{\mu }\right)
u\left( p\right) =\frac{p^{\mu }}{mc}\overline{u}\left( p\right) u\left(
p\right) \label{rd12d}
\end{equation}%
for an arbitrary normalization of the spinor $\phi $ in (\ref{rd6}) (see
\cite{Moskalev} for more details).

\subsection{Expectation Values of Relativistic Spin Operator}

Let%
\begin{equation}
\ \left\langle \boldsymbol{\widehat{\boldsymbol{S}}}\right\rangle =\frac{%
u^{\dag }\left( \frac{1}{2}\mathbf{\Sigma }\right) u}{u^{\dag }u},\qquad
\left\langle \boldsymbol{\widehat{\boldsymbol{s}}}\right\rangle =\frac{\phi
^{\dag }\left( \frac{1}{2}\mathbf{\sigma }\right) \phi }{\phi ^{\dag }\phi }
\label{sr1}
\end{equation}%
be expectation values for the relativistic, when $\boldsymbol{p\neq }\mathbf{%
0},$ and nonrelativistic, if $\boldsymbol{p=}\mathbf{0},$ spin operators for
a Dirac particle, respectively, \cite{Moskalev}. Then
{\footnote{This result follows also from (\ref{rd9rd})--(\ref{rd10})}}
\begin{equation}
\ \left\langle \boldsymbol{\widehat{\boldsymbol{S}}}\right\rangle =\frac{%
mc^{2}}{E}\left\langle \boldsymbol{\widehat{\boldsymbol{s}}}\right\rangle +%
\frac{c^{2}\boldsymbol{p}}{E\left( E+mc^{2}\right) }\left( \boldsymbol{%
p\cdot }\left\langle \boldsymbol{\widehat{\boldsymbol{s}}}\right\rangle
\right) . \label{sr2}
\end{equation}%
Indeed, for bi-spinor (\ref{rd6}), up to a constant,%
\begin{eqnarray}
u^{\dag }\left( \mathbf{\Sigma }\right) u &=&\left( \phi ^{\dag },\phi
^{\dag }\dfrac{c\left( \boldsymbol{\sigma \cdot p}\right) }{mc^{2}+E}\right)
\left(
\begin{array}{cc}
\mathbf{\sigma } & \mathbf{0} \\
\mathbf{0} & \mathbf{\sigma }%
\end{array}%
\right) \left(
\begin{array}{c}
\phi \\
\dfrac{c\left( \boldsymbol{\sigma \cdot p}\right) }{mc^{2}+E}\phi%
\end{array}%
\right) \label{sr3} \\
&=&\left( \phi ^{\dag },\phi ^{\dag }\dfrac{c\left( \boldsymbol{\sigma \cdot
p}\right) }{mc^{2}+E}\right) \left(
\begin{array}{c}
\mathbf{\sigma }\phi \\
\mathbf{\sigma }\dfrac{c\left( \boldsymbol{\sigma \cdot p}\right) }{mc^{2}+E}%
\phi%
\end{array}%
\right) \notag \\
&=&\phi ^{\dag }\mathbf{\sigma }\phi +\frac{c^{2}}{\left( E+mc^{2}\right)
^{2}}\phi ^{\dag }\left[ \left( \boldsymbol{\sigma \cdot p}\right) \mathbf{%
\sigma }\left( \boldsymbol{\sigma \cdot p}\right) \right] \phi . \notag
\end{eqnarray}%
In view of (\ref{rd11}), one gets%
\begin{equation}
u^{\dag }\left( \mathbf{\Sigma }\right) u=\frac{2mc^{2}}{E+mc^{2}}\left[
\left( \phi ^{\dag }\mathbf{\sigma }\phi \right) +\frac{\boldsymbol{p}}{%
m\left( E+mc^{2}\right) }\boldsymbol{p\cdot }\left( \phi ^{\dag }\mathbf{%
\sigma }\phi \right) \right] . \label{sr4}
\end{equation}%
In a similar fashion, up to a constant,
\begin{equation}
u^{\dag }u=\frac{2E}{E+mc^{2}}\left( \phi ^{\dag }\phi \right) . \label{sr5}
\end{equation}%
As a result, we obtain (\ref{sr2}).

In a special case, when the $z$-axis is directed towars $\boldsymbol{p},$
one gets%
\begin{equation}
\ \left\langle \widehat{S_{x}}\right\rangle =\frac{mc^{2}}{E}\left\langle
\widehat{s_{x}}\right\rangle ,\qquad \ \left\langle \widehat{S_{y}}%
\right\rangle =\frac{mc^{2}}{E}\left\langle \widehat{s_{y}}\right\rangle
,\qquad \ \left\langle \widehat{S_{z}}\right\rangle =\left\langle \widehat{%
s_{z}}\right\rangle \label{sr6}
\end{equation}%
(see \cite{Moskalev} for more details).

\subsection{Relativistic Helicity States}

When the spinor $\phi $ (respectively, $\chi $) corresponds to the
nonrelativistic helicity states (\ref{s7})--(\ref{s8b}) with the particular
polarization vector $\boldsymbol{\boldsymbol{n}}=\boldsymbol{\boldsymbol{p/}}%
\left\vert \boldsymbol{\boldsymbol{p}}\right\vert $ (in the frame of
reference where the particle is at rest), the bi-spinors (\ref{s13}) become
the common eigenfunctions of the commuting Hamiltonian and helicity
operator: $\left[ \widehat{H},\ \Lambda \right] =0.$ For example, up to a
normalization,%
\begin{eqnarray}
\left( \boldsymbol{\Sigma \cdot \boldsymbol{p}}\right) v\left( p\right)
&=&C\left(
\begin{array}{cc}
\boldsymbol{\sigma \cdot p} & \boldsymbol{0} \\
\boldsymbol{0} & \boldsymbol{\sigma \cdot p}%
\end{array}%
\right) \left(
\begin{array}{c}
\phi \\
\dfrac{c\left( \boldsymbol{\sigma \cdot p}\right) }{mc^{2}+E}\phi%
\end{array}%
\right) \label{s14} \\
&=&C\left(
\begin{array}{c}
\left( \boldsymbol{\sigma \cdot p}\right) \phi \\
\left( \boldsymbol{\sigma \cdot p}\right) \dfrac{c\left( \boldsymbol{\sigma
\cdot p}\right) }{mc^{2}+E}\phi%
\end{array}%
\right) =2\lambda \left\vert \boldsymbol{\boldsymbol{p}}\right\vert C\left(
\begin{array}{c}
\phi \\
\dfrac{c\left( \boldsymbol{\sigma \cdot p}\right) }{mc^{2}+E}\phi%
\end{array}%
\right) , \notag
\end{eqnarray}%
or%
\begin{equation}
\widehat{\Lambda }v\left( p\right) =\lambda v\left( p\right) \label{s15}
\end{equation}%
for our both helicity states (\ref{s13}).

Applying the block matrix multiplication rule (\ref{A1}), one gets:
\begin{eqnarray}
&&\left(
\begin{array}{cc}
\left( mc^{2}+E\right) \Phi & c\left( \boldsymbol{\sigma \cdot p}\right) \Phi
\\
c\left( \boldsymbol{\sigma \cdot p}\right) \Phi & -\left( mc^{2}+E\right)
\Phi%
\end{array}%
\right) ^{\dag }\left(
\begin{array}{cc}
\left( mc^{2}+E\right) \Phi & c\left( \boldsymbol{\sigma \cdot p}\right) \Phi
\\
c\left( \boldsymbol{\sigma \cdot p}\right) \Phi & -\left( mc^{2}+E\right)
\Phi%
\end{array}%
\right) \label{s16} \\
&&\quad =\left(
\begin{array}{cc}
\left( mc^{2}+E\right) \Phi ^{\dag } & c\Phi ^{\dag }\left( \boldsymbol{%
\sigma \cdot p}\right) \\
c\Phi ^{\dag }\left( \boldsymbol{\sigma \cdot p}\right) & -\left(
mc^{2}+E\right) \Phi ^{\dag }%
\end{array}%
\right) \left(
\begin{array}{cc}
\left( mc^{2}+E\right) \Phi & c\left( \boldsymbol{\sigma \cdot p}\right) \Phi
\\
c\left( \boldsymbol{\sigma \cdot p}\right) \Phi & -\left( mc^{2}+E\right)
\Phi%
\end{array}%
\right) \notag \\
&&\qquad \qquad \qquad \qquad \qquad \qquad \qquad \qquad =\left( \left(
mc^{2}+E\right) ^{2}+c^{2}\boldsymbol{p}^{2}\right) \left(
\begin{array}{cc}
\mathbf{1} & \boldsymbol{0} \\
\boldsymbol{0} & \boldsymbol{1}%
\end{array}%
\right) \notag
\end{eqnarray}%
as an extension of our identity (\ref{ma2}). Once again, all relativistic
helicity states can be unified as the columns in the following $4\times 4$
unitary matrix:%
\begin{eqnarray}
V &=&\left( v_{+}^{\left( 1/2\right) },v_{+}^{\left( -1/2\right)
},v_{-}^{\left( 1/2\right) },v_{-}^{\left( -1/2\right) }\right) \label{s17}
\\
&=&\frac{1}{\sqrt{\left( mc^{2}+R\right) ^{2}+c^{2}\boldsymbol{p}^{2}}}%
\left(
\begin{array}{cc}
\left( mc^{2}+R\right) \Phi & c\left( \boldsymbol{\sigma \cdot p}\right) \Phi
\\
c\left( \boldsymbol{\sigma \cdot p}\right) \Phi & -\left( mc^{2}+R\right)
\Phi%
\end{array}%
\right) \notag \\
&=&\sqrt{\frac{mc^{2}+R}{2R}}\left(
\begin{array}{cc}
\Phi & \dfrac{c\left\vert \boldsymbol{\boldsymbol{p}}\right\vert }{mc^{2}+R}%
\widetilde{\Phi } \\
\dfrac{c\left\vert \boldsymbol{\boldsymbol{p}}\right\vert }{mc^{2}+R}%
\widetilde{\Phi } & -\Phi%
\end{array}%
\right) ,\qquad V^{-1}= \gamma^0 V^{\dag } \gamma^0 \notag
\end{eqnarray}%
with the help of (\ref{s8h}), that is similar to (\ref{ma6}):%
\begin{eqnarray}
V &=&\left( v_{+}^{\left( 1/2\right) },v_{+}^{\left( -1/2\right)
},v_{-}^{\left( 1/2\right) },v_{-}^{\left( -1/2\right) }\right) \label{s18}
\\
&=&\sqrt{\frac{mc^{2}+R}{2R}}\left(
\begin{array}{cc}
\left( \phi ^{\left( 1/2\right) },\ \phi ^{\left( -1/2\right) }\right) &
\dfrac{c\left\vert \boldsymbol{\boldsymbol{p}}\right\vert }{mc^{2}+R}\left(
\phi ^{\left( 1/2\right) },\ -\phi ^{\left( -1/2\right) }\right) \\
\dfrac{c\left\vert \boldsymbol{p}\right\vert }{mc^{2}+R}\left( \phi ^{\left(
1/2\right) },\ -\phi ^{\left( -1/2\right) }\right) & -\left( \phi ^{\left(
1/2\right) },\ \phi ^{\left( -1/2\right) }\right)%
\end{array}%
\right) . \notag
\end{eqnarray}%
Here, the first two columns correspond to the positive energy eigenvalues $%
E=+R=\sqrt{c^{2}\boldsymbol{p}^{2}+m^{2}c^{4}}$ and the helicities $\lambda
=\pm 1/2,$ respec\-tively. Whereas, the last two columns correspond to the
negative energy eigenvalues, given by $E=-R=-\sqrt{c^{2}\boldsymbol{p}%
^{2}+m^{2}c^{4}},$ and the helicities $\lambda =\pm 1/2,$ respec\-tively.
One should verify these facts by direct substitutions, similar to (\ref{ma7a}%
)--(\ref{ma7b}).

For the Hamiltonian $4\times 4$ matrix:%
\begin{equation}
H=c\boldsymbol{\alpha \cdot \boldsymbol{p}}+mc^{2}\beta , \label{s19}
\end{equation}%
one gets%
\begin{equation}
H^{2}=\left( c\boldsymbol{\alpha \cdot \boldsymbol{p}}+mc^{2}\beta \right)
^{2}=R^{2}=c^{2}\boldsymbol{p}^{2}+m^{2}c^{4} \label{s20}
\end{equation}%
due to the familiar anticommutators:%
\begin{equation}
\alpha _{r}\alpha _{s}+\alpha _{s}\alpha _{r}=2\delta _{rs},\qquad \alpha
_{r}\beta +\beta \alpha _{r}=0\qquad \left( r,s=1,2,3\right) . \label{s21}
\end{equation}%
As a result, we arrive at the following matrix version of the eigenvalue
problem under consideration:%
\begin{equation}
HV=R\widetilde{V},\qquad H\widetilde{V}=RV, \label{s22}
\end{equation}%
where, by definition,%
\begin{equation}
\widetilde{V}=\left( v_{+}^{\left( 1/2\right) },v_{+}^{\left( -1/2\right)
},-v_{-}^{\left( 1/2\right) },-v_{-}^{\left( -1/2\right) }\right), \qquad  {\widetilde{V}}^{-1}={\gamma}^0 {\widetilde{V}}^{\dag} {\gamma}^0 .
\label{s23}
\end{equation}%
One can obtain the following decompositions%
\begin{equation}
H=R\widetilde{V}\ V^{-1}=RV\ \widetilde{V}^{-1 } \label{s24}
\end{equation}%
similar to our factorization (\ref{s8e}) of the nonrelativistic helicity
matrix. (These results are verified in the Mathematica file.)

\subsection{Convenient Parametrization}

Following \cite{Flugge}, with somewhat different details, let us also
discuss the Dirac plane waves of positive and negative helicities $\lambda
=\pm 1/2,$ but of positive energy only, in a slightly different notation.
With%
\begin{equation}
\psi =ve^{ \iun \left( \boldsymbol{k\cdot r}-\omega t\right) } \label{h1}
\end{equation}%
and using the abbreviations%
\begin{equation}
k\eta =\frac{\omega }{c}-\frac{mc}{\hbar },\qquad \frac{k}{\eta }=\frac{%
\omega }{c}+\frac{mc}{\hbar } \label{h2}
\end{equation}%
one can express the magnitude of particle linear momentum and its kinetic
energy in terms of a suitable parameter $\eta :$%
\begin{equation}
\left\vert \boldsymbol{\boldsymbol{p}}\right\vert =\hbar k=mc\frac{2\eta }{%
1-\eta ^{2}},\qquad E=\hbar \omega =mc^{2}\frac{1+\eta ^{2}}{1-\eta ^{2}}.
\label{h3}
\end{equation}%
Once again, we use two polar angles $\theta $ and $\varphi $ in the
direction of vector $\boldsymbol{k}=\boldsymbol{p}/\hbar :$%
\begin{equation}
k_{1}\pm  \iun k_{2}=k\sin \theta \ e^{\pm  \iun \varphi },\qquad k_{3}=k\cos \theta .
\label{h4}
\end{equation}%
The eigenfunctions of the helicity operator are given by (\ref{s8a})--(\ref%
{s8b}) and (\ref{s18}) \cite{Flugge}:%
\begin{equation}
v_{+}^{\left( 1/2\right) }=\frac{1}{\sqrt{V\left( 1+\eta ^{2}\right) }}%
\left(
\begin{array}{c}
\cos \dfrac{\theta }{2}\exp \left( - \iun \dfrac{\varphi }{2}\right) \\
\sin \dfrac{\theta }{2}\exp \left(  \iun \dfrac{\varphi }{2}\right) \\
\eta \cos \dfrac{\theta }{2}\exp \left( - \iun \dfrac{\varphi }{2}\right) \\
\eta \sin \dfrac{\theta }{2}\exp \left(  \iun \dfrac{\varphi }{2}\right)%
\end{array}%
\right) \label{h5}
\end{equation}%
for $\lambda =+1/2$ and%
\begin{equation}
v_{+}^{\left( -1/2\right) }=\frac{1}{\sqrt{V\left( 1+\eta ^{2}\right) }}%
\left(
\begin{array}{c}
\sin \dfrac{\theta }{2}\exp \left( - \iun \dfrac{\varphi }{2}\right) \\
-\cos \dfrac{\theta }{2}\exp \left(  \iun \dfrac{\varphi }{2}\right) \\
-\eta \sin \dfrac{\theta }{2}\exp \left( - \iun \dfrac{\varphi }{2}\right) \\
\eta \cos \dfrac{\theta }{2}\exp \left(  \iun \dfrac{\varphi }{2}\right)%
\end{array}%
\right) \label{h6}
\end{equation}%
for $\lambda =-1/2.$ [One should compare these bi-spinors with the first two
columns in our $4\times 4$ helicity states matrix (\ref{s18}), where, in
turn, the last two columns correspond to the negative energy eigenvalues.]
Here, we use the standard normalization%
\begin{equation}
\int_{V}\overline{\psi }\gamma ^{0}\psi \ dv=\int_{V}\psi ^{\dag }\psi \
dv=\int_{V}u^{\dag }u\ dv=1. \label{h7}
\end{equation}%
It should be noted that this normalization is Lorentz-invariant, the
integral being proportional to the total electric charge inside the volume $%
V.$ (See \cite{Flugge} and \cite{Moskalev} for more details.)

In the nonrelativistic limit, when $\eta \ll 1,$ the last two components in
the above bi-spinors may be neglected and we arrive at the two-component
Pauli spin theory. The case of negative energy eigenvalues will be discussed later.

\section{The Polarization Density Matrices\/}

\subsection{Positive Energy Eigenvalues}

For a Dirac particle, with a given linear momentum $\boldsymbol{\boldsymbol{p%
}},$ positive energy $E,$ and a certain polarization $\lambda ,$ the wave
function (\ref{rd4}) is defined by the corresponding bi-spinor $u^{\left(
\lambda \right) }\left( p\right) ,$ which satisfies the following equations
in the momentum representation \cite{Moskalev}:%
\begin{equation}
\left( \widehat{p}-mc\right) u^{\left( \lambda \right) }\left( p\right)
=0,\qquad \overline{u}^{\left( \lambda \right) }\left( p\right) \left(
\widehat{p}-mc\right) =0, \label{po1}
\end{equation}%
where, by definition, $\widehat{p}=\gamma ^{\mu }p_{\mu }=\gamma ^{0}\left(
E/c\right) -\left( \boldsymbol{\gamma \cdot p}\right) $ and $\overline{u}%
^{\left( \lambda \right) }\left( p\right) =\left( u^{\left( \lambda \right)
}\left( p\right) \right) ^{\dag }\gamma ^{0}.$ Here, we choose the following
normalization%
\begin{equation}
\overline{u}^{\left( \lambda \right) }\left( p\right) u^{\left( \lambda
\right) }\left( p\right) =2mc. \label{po2}
\end{equation}%
By (\ref{rd6})--(\ref{rd8}), those bi-spinor solutions have the form%
\begin{equation}
u^{\left( \lambda \right) }\left( p\right) =\sqrt{E+mc^{2}}\left(
\begin{array}{c}
\phi ^{\left( \lambda \right) } \\
\dfrac{c\left( \boldsymbol{\sigma \cdot p}\right) }{mc^{2}+E}\phi ^{\left(
\lambda \right) }%
\end{array}%
\right) ,\quad \psi _{+}\left( x\right) =u^{\left( \lambda \right) }\left(
p\right) e^{ \iun \left( \boldsymbol{p\cdot r}-Et\right) /\hbar } \label{po3}
\end{equation}%
Here $E=+R=\sqrt{c^{2}\boldsymbol{p}^{2}+m^{2}c^{4}}$ and the spinor $\phi
^{\left( \lambda \right) }$ satisfies equations (\ref{rd9}). (With a given
linear momentum $\boldsymbol{\boldsymbol{p}}$ and positive energy $E,$ there
are two possible polarization states of the particle; for example, the
states with the given helicity $\lambda =\pm 1/2$ or the states with the
projection of the spin on the third axis $S_{3}=\pm 1/2,$ in the frame of
reference when the particle is at rest; see \cite{BerLifPit}, \cite{Moskalev}
for more details.)

For evaluation of the relativistic scattering amplitudes for the spin $1/2$
particle, the following bilinear form are important \cite{BerLifPit}, \cite%
{Moskalev}:%
\begin{equation}
\sum_{\lambda }u_{\alpha }^{\left( \lambda \right) }\left( p\right)
\overline{u}_{\beta }^{\left( \lambda \right) }\left( p\right) :=\left[
\Lambda _{+}\left( p\right) \right] _{\alpha \beta }\quad \quad \left(
\alpha ,\beta =1,2,3,4\right) . \label{po4}
\end{equation}%
Here, $\Lambda _{+}\left( p\right) $ is a certain $4\times 4$ matrix, the
symbol $+$ denotes the states with the positive energy eigenvalues, and the
summation is performed over both polarizations.

From equations (\ref{po1}) and normalization (\ref{po2}) one gets%
\begin{eqnarray}
\left( \widehat{p}-mc\right) \Lambda _{+}\left( p\right) &=&0,\qquad \Lambda
_{+}\left( p\right) \left( \widehat{p}-mc\right) =0, \label{po5} \\
\text{Tr\ }\Lambda _{+}\left( p\right) &=&2mc \notag
\end{eqnarray}%
and%
\begin{equation}
\Lambda _{+}\left( p\right) =mc+\widehat{p}, \label{po6}
\end{equation}%
which can be explicitly verified from the definition (\ref{po4}) with the
help of the following identity%
\begin{equation}
\sum_{\lambda =-1/2}^{1/2}\phi _{r}^{\left( \lambda \right) }\left(
\boldsymbol{\boldsymbol{n}}\right) \left[ \phi _{s}^{\left( \lambda \right)
}\left( \boldsymbol{\boldsymbol{n}}\right) \right] ^{\ast }=\delta
_{rs}\qquad \left( r,s=1,2\right) \label{po7}
\end{equation}%
for the spinors (\ref{s8a})--(\ref{s8b}).

\subsection{Negative Energy Eigenvalues}

In addition to (\ref{rd4}), the Dirac equation (\ref{rd1}) has also the
following solutions in covariant form%
\begin{equation}
\psi _{-}\left( x\right) =v\left( p\right) e^{ \iun \left( p_{\mu }x^{\mu
}\right) /\hbar },\qquad \left( \widehat{p}+mc\right) v\left( p\right)
=\left( \gamma ^{\mu }p_{\mu }+mc\right) v\left( p\right) =0, \label{po7a}
\end{equation}%
which can be verified by a direct substitution \cite{Moskalev}. On the
second thought, for the states with a negative energy eigenvalues, when $%
E=-R=-\sqrt{c^{2}\boldsymbol{p}^{2}+m^{2}c^{4}}$ and $\boldsymbol{p}%
\rightarrow -$ $\boldsymbol{p},$ say, in the second relation (\ref{s13}),
with a certain polarization $\lambda ,$ once again, the corresponding
bi-spinor solutions can be written as follows%
\begin{equation}
v^{\left( \lambda \right) }\left( p\right) =\sqrt{R+mc^{2}}\left(
\begin{array}{c}
\dfrac{c\left( \boldsymbol{\sigma \cdot p}\right) }{R+mc^{2}}\phi ^{\left(
\lambda \right) } \\
\phi ^{\left( \lambda \right) }%
\end{array}%
\right) ,\quad \psi _{-}\left( x\right) =v^{\left( \lambda \right) }\left(
p\right) e^{ \iun \left( Rt-\boldsymbol{p\cdot r}\right) /\hbar }. \label{po8}
\end{equation}%
Here,%
\begin{equation}
\overline{v}^{\left( \lambda \right) }\left( p\right) v^{\left( \lambda
\right) }\left( p\right) =-2mc \label{po8a}
\end{equation}%
(cf. \cite{AkhBer}, \cite{BerLifPit}, \cite{BBQED}), and%
\begin{equation}
\left( \widehat{p}+mc\right) v^{\left( \lambda \right) }\left( p\right)
=0,\qquad \overline{v}^{\left( \lambda \right) }\left( p\right) \left(
\widehat{p}+mc\right) =0. \label{po9}
\end{equation}%
Moreover, the bi-spinors $u^{\left( \lambda \right) }\left( p\right) $ and $%
v^{\left( \lambda \right) }\left( p\right) $ satisfy the orthogonality
condition%
\begin{equation}
\overline{u}^{\left( \lambda \right) }\left( p\right) v^{\left( \lambda
^{\prime }\right) }\left( p\right) =0\qquad \left( \lambda \neq \lambda
^{\prime }\right) . \label{po9a}
\end{equation}

In a similar fashion, for the negative energy eigenvalues, we introduce%
\begin{equation}
\sum_{\lambda }v_{\alpha }^{\left( \lambda \right) }\left( p\right)
\overline{v}_{\beta }^{\left( \lambda \right) }\left( p\right) :=-\left[
\Lambda _{-}\left( p\right) \right] _{\alpha \beta }\quad \quad \left(
\alpha ,\beta =1,2,3,4\right) \label{po10}
\end{equation}%
and derive%
\begin{equation}
\Lambda _{-}\left( p\right) =mc-\widehat{p}. \label{po11}
\end{equation}

The matrix operators $\Lambda _{+}\left( p\right) $\ and $\Lambda _{-}\left(
p\right) $ obey the following properties%
\begin{eqnarray}
&&\qquad \Lambda _{+}\left( p\right) +\Lambda _{-}\left( p\right) =2mc,
\label{po12} \\
&&\Lambda _{+}\left( p\right) \Lambda _{-}\left( p\right) =\Lambda
_{-}\left( p\right) \Lambda _{+}\left( p\right) =0, \notag \\
&&\Lambda _{+}^{2}\left( p\right) =2mc\ \Lambda _{+}\left( p\right) ,\quad
\Lambda _{-}^{2}\left( p\right) =2mc\ \Lambda _{-}\left( p\right) , \notag
\end{eqnarray}%
which can be easily verified with the help of the relativistic invariant $%
\left( \widehat{p}\right) ^{2}=p^{2}=m^{2}c^{2}.$ Indeed, in compact form,%
\begin{eqnarray}
\left( \widehat{p}\right) ^{2} &=&\left( \gamma ^{\mu }p_{\mu }\right)
\left( \gamma ^{\nu }p_{\nu }\right) =\left( \gamma ^{\mu }\gamma ^{\nu
}\right) \left( p_{\mu }p_{\nu }\right) \label{po13} \\
&=&\frac{1}{2}\left( \gamma ^{\mu }\gamma ^{\nu }+\gamma ^{\nu }\gamma ^{\mu
}\right) \left( p_{\mu }p_{\nu }\right) \notag \\
&&+\frac{1}{2}\left( \gamma ^{\mu }\gamma ^{\nu }-\gamma ^{\nu }\gamma ^{\mu
}\right) \left( p_{\mu }p_{\nu }\right) \notag \\
&=&g^{\mu \nu }p_{\mu }p_{\nu }=p_{\mu }p^{\mu }=p^{2} \notag
\end{eqnarray}%
by (\ref{rd3a}). The last two relations in (\ref{po12}), up to a
normalization, define the so-called projection operators to the states with
positive and negative energy eigenvalues, respectively \cite{Moskalev}.
Moreover, two the last but one equations are similar to our matrix product (%
\ref{ma1}) for the relativistic spectrum.

\subsection{Polarization Density Matrix for Pure States}

Moreover, for Dirac's particle, with a given polarization $\lambda =\pm 1/2,$
one can show that \cite{AkhBer}, \cite{Moskalev}:%
\begin{eqnarray}
u_{\alpha }^{\left( \lambda \right) }\left( p\right) \overline{u}_{\beta
}^{\left( \lambda \right) }\left( p\right) &:&=\left[ \rho _{+}\left(
p\right) \right] _{\alpha \beta },\qquad \rho _{+}\left( p\right) =\frac{1}{2%
}\left( mc+\widehat{p}\right) \left( I-\gamma _{5}\widehat{a}\right) ;
\label{po14} \\
v_{\alpha }^{\left( \lambda \right) }\left( p\right) \overline{v}_{\beta
}^{\left( \lambda \right) }\left( p\right) &:&=-\left[ \rho _{-}\left(
p\right) \right] _{\alpha \beta },\qquad \rho _{-}\left( p\right) =\frac{1}{2%
}\left( mc-\widehat{p}\right) \left( I-\gamma _{5}\widehat{a}\right)
\label{po14aa}
\end{eqnarray}%
(no summation over $\lambda $ in both equations), where $\widehat{a}=$ $%
\gamma ^{\mu }a_{\mu }$ and $a^{\mu }$ is the polarization four vector (\ref%
{rd10}). As is worth noting, summation over two possible polarizations, in
each of the above equations (\ref{po14})--(\ref{po14aa}), with different
signs of $\pm \boldsymbol{\boldsymbol{n}}$ and, therefore, in all components
of $a^{\mu },$ results in (\ref{po6}) and (\ref{po11}), as expected. (More
details on the polarization density matrices can be found in \cite{AkhBer},
\cite{BerLifPit}, and \cite{Moskalev}.)

These relations can be verified by a direct matrix multiplication. Indeed,
say for the bi-spinor (\ref{h5}) with $\lambda =1/2$ and positive energy
eigenvalues, one gets, for both polarizations $\lambda =\pm 1/2,$ that
\begin{equation}
mc+\widehat{p}=\frac{2mc}{1-\eta ^{2}}\left(
\begin{array}{cccc}
1 & 0 & -\eta \cos \theta & -\eta \sin \theta \ e^{- \iun \varphi } \\
0 & 1 & -\eta \sin \theta \ e^{ \iun \varphi } & \eta \cos \theta \\
\eta \cos \theta & \eta \sin \theta \ e^{- \iun \varphi } & -\eta ^{2} & 0 \\
\eta \sin \theta \ e^{ \iun \varphi } & -\eta \cos \theta & 0 & -\eta ^{2}%
\end{array}%
\right) . \label{po14a}
\end{equation}%
Moreover, for $\lambda =1/2,$
\begin{equation}
I-\gamma _{5}\widehat{a}=\left(
\begin{array}{cc}
\mathbf{1}+\boldsymbol{\sigma \cdot a} & -a_{0}\mathbf{1} \\
a_{0}\mathbf{1} & \mathbf{1}-\boldsymbol{\sigma \cdot a}%
\end{array}%
\right) , \label{po15}
\end{equation}%
where%
\begin{eqnarray}
\mathbf{1}+\boldsymbol{\sigma \cdot a} &=&\frac{2}{1-\eta ^{2}}\left(
\begin{array}{cc}
1-\left( 1+\eta ^{2}\right) \sin ^{2}\dfrac{\theta }{2} & \left( 1+\eta
^{2}\right) \sin \dfrac{\theta }{2}\cos \dfrac{\theta }{2}\ e^{-
	\iun \varphi }
\\
\left( 1+\eta ^{2}\right) \sin \dfrac{\theta }{2}\cos \dfrac{\theta }{2}\
e^{ \iun \varphi } & 1-\left( 1+\eta ^{2}\right) \cos ^{2}\dfrac{\theta }{2}%
\end{array}%
\right) , \label{po16} \\
\mathbf{1}-\boldsymbol{\sigma \cdot a} &=&\frac{2}{1-\eta ^{2}}\left(
\begin{array}{cc}
1-\left( 1+\eta ^{2}\right) \cos ^{2}\dfrac{\theta }{2} & -\left( 1+\eta
^{2}\right) \sin \dfrac{\theta }{2}\cos \dfrac{\theta }{2}\ e^{-
	\iun \varphi }
\\
-\left( 1+\eta ^{2}\right) \sin \dfrac{\theta }{2}\cos \dfrac{\theta }{2}\
e^{ \iun \varphi } & 1-\left( 1+\eta ^{2}\right) \sin ^{2}\dfrac{\theta }{2}%
\end{array}%
\right) \label{po16a}
\end{eqnarray}%
by (\ref{rd10}). As a result,%
%
%
\begin{eqnarray}
&&\left( \frac{2}{1-\eta ^{2}}\right) ^{-1}\left( I-\gamma _{5}\widehat{a}%
\right)  \label{po17} \\
&=&\left(
\begin{array}{cccc}
\cos ^{2}\dfrac{\theta }{2}-\eta ^{2}\sin ^{2}\dfrac{\theta }{2} & \dfrac{1}{%
2}\left( 1+\eta ^{2}\right) \sin \theta \ e^{- \iun \varphi } & -\eta & 0 \\
\dfrac{1}{2}\left( 1+\eta ^{2}\right) \sin \theta \ e^{ \iun \varphi } & \sin ^{2}%
\dfrac{\theta }{2}-\eta ^{2}\cos ^{2}\dfrac{\theta }{2} & 0 & -\eta \\
\eta & 0 & \sin ^{2}\dfrac{\theta }{2}-\eta ^{2}\cos ^{2}\dfrac{\theta }{2}
& -\dfrac{1}{2}\left( 1+\eta ^{2}\right) \sin \theta \ e^{- \iun \varphi } \\
0 & \eta & -\dfrac{1}{2}\left( 1+\eta ^{2}\right) \sin \theta \
	e^{ \iun \varphi }
& \cos ^{2}\dfrac{\theta }{2}-\eta ^{2}\sin ^{2}\dfrac{\theta }{2}%
\end{array}%
\right) .  \notag
\end{eqnarray}%
Finally, by multiplication of the matrices (\ref{po14a}) and (\ref{po17}),
we obtain%
\begin{equation}
\left( 1-\eta ^{2}\right) \left(
\begin{array}{cccc}
\cos ^{2}\dfrac{\theta }{2} & \frac{1}{2}\sin \theta \ e^{- \iun \varphi } &
-\eta \cos ^{2}\dfrac{\theta }{2} & -\frac{1}{2}\eta \sin \theta \
e^{- \iun \varphi } \\
\frac{1}{2}\sin \theta \ e^{ \iun \varphi } & \sin ^{2}\dfrac{\theta }{2} & -%
\frac{1}{2}\eta \sin \theta \ e^{ \iun \varphi } & -\eta \sin ^{2}\dfrac{\theta }{%
2} \\
\eta \cos ^{2}\dfrac{\theta }{2} & \frac{1}{2}\eta \sin \theta \
e^{- \iun \varphi } & -\eta ^{2}\cos ^{2}\dfrac{\theta }{2} & -\frac{1}{2}\eta
^{2}\sin \theta \ e^{- \iun \varphi } \\
\frac{1}{2}\eta \sin \theta \ e^{ \iun \varphi } & \eta \sin ^{2}\dfrac{\theta }{2%
} & -\frac{1}{2}\eta ^{2}\sin \theta \ e^{ \iun \varphi } & -\eta ^{2}\sin ^{2}%
\dfrac{\theta }{2}%
\end{array}%
\right) . \label{po18}
\end{equation}%
As is easy to verify, this matrix is equal to the left hand side, namely the
tensor product $\left[ \rho _{+}\left( p\right) \right] _{\alpha \beta },$
of the relation (\ref{po14}) for the bi-spinor (\ref{h5}), up to the
required constant. The case $\lambda =-1/2$ and $E>0$ can be verified in a
similar fashion, with the help of a computer algebra system (see our Mathematica file).

For the negative energy eigenvalues, one can present the corresponding
bi-spinors (\ref{po8}), as follows
\begin{equation}
v_{-}^{\left( \lambda =-1/2\right) }=\frac{1}{\sqrt{V\left( 1+\eta
^{2}\right) }}\left(
\begin{array}{c}
\eta \cos \dfrac{\theta }{2}\exp \left( - \iun \dfrac{\varphi }{2}\right) \\
\eta \sin \dfrac{\theta }{2}\exp \left(  \iun \dfrac{\varphi }{2}\right) \\
\cos \dfrac{\theta }{2}\exp \left( - \iun \dfrac{\varphi }{2}\right) \\
\sin \dfrac{\theta }{2}\exp \left(  \iun \dfrac{\varphi }{2}\right)
\end{array}%
\right) , \label{po19a}
\end{equation}%
and%
\begin{equation}
v_{-}^{\left( \lambda =1/2\right) }=\frac{1}{\sqrt{V\left( 1+\eta
^{2}\right) }}\left(
\begin{array}{c}
\eta \sin \dfrac{\theta }{2}\exp \left( - \iun \dfrac{\varphi }{2}\right) \\
-\eta \cos \dfrac{\theta }{2}\exp \left(  \iun \dfrac{\varphi }{2}\right) \\
-\sin \dfrac{\theta }{2}\exp \left( - \iun \dfrac{\varphi }{2}\right) \\
\cos \dfrac{\theta }{2}\exp \left(  \iun \dfrac{\varphi }{2}\right)
\end{array}%
\right) , \label{po19b}
\end{equation}%
say, in the normalization (\ref{h7}). [It should be noted that the
substitution $E\rightarrow -E$ and $\boldsymbol{p}\rightarrow -$ $%
\boldsymbol{p},$ in (\ref{po7a})--(\ref{po8}), implies that $\lambda
\rightarrow -\lambda .]$ Here and in (\ref{h5})--(\ref{h6}), one has to
replace%
\begin{equation}
\frac{1}{\sqrt{V\left( 1+\eta ^{2}\right) }}\rightarrow \sqrt{\frac{2mc}{%
1-\eta ^{2}}} \label{po19ab}
\end{equation}%
in the case of the relativistically invariant normalization (\ref{po8a}),
which we are using in this section. Once again, for both polarizations $%
\lambda =\pm 1/2,$ one gets
\begin{equation}
mc-\widehat{p}=-\frac{2mc}{1-\eta ^{2}}\left(
\begin{array}{cccc}
\eta ^{2} & 0 & -\eta \cos \theta & -\eta \sin \theta \ e^{- \iun \varphi } \\
0 & \eta ^{2} & -\eta \sin \theta \ e^{ \iun \varphi } & \eta \cos \theta \\
\eta \cos \theta & \eta \sin \theta \ e^{- \iun \varphi } & -1 & 0 \\
\eta \sin \theta \ e^{ \iun \varphi } & -\eta \cos \theta & 0 & -1%
\end{array}%
\right) . \label{po20}
\end{equation}%
Let us choose $\lambda =-1/2,$ when one can use (\ref{rd10}), once again,
but with $E\rightarrow \left\vert E\right\vert =R=\sqrt{c^{2}\boldsymbol{p}%
^{2}+m^{2}c^{4}}$ and $\boldsymbol{n}\rightarrow -\boldsymbol{n}.$ As a
result,
%
\begin{eqnarray}
&&\left( \frac{2}{1-\eta ^{2}}\right) ^{-1}\left( I+\gamma _{5}\widehat{a}%
\right)  \label{po21} \\
&=&\left(
\begin{array}{cccc}
\sin ^{2}\dfrac{\theta }{2}-\eta ^{2}\cos ^{2}\dfrac{\theta }{2} & -\dfrac{1%
}{2}\left( 1+\eta ^{2}\right) \sin \theta \ e^{- \iun \varphi } & \eta & 0 \\
-\dfrac{1}{2}\left( 1+\eta ^{2}\right) \sin \theta \ e^{ \iun \varphi } & \cos
^{2}\dfrac{\theta }{2}-\eta ^{2}\sin ^{2}\dfrac{\theta }{2} & 0 & \eta \\
-\eta & 0 & \cos ^{2}\dfrac{\theta }{2}-\eta ^{2}\sin ^{2}\dfrac{\theta }{2}
& \dfrac{1}{2}\left( 1+\eta ^{2}\right) \sin \theta \ e^{- \iun \varphi } \\
0 & -\eta & \dfrac{1}{2}\left( 1+\eta ^{2}\right) \sin \theta \
	e^{ \iun \varphi }
& \sin ^{2}\dfrac{\theta }{2}-\eta ^{2}\cos ^{2}\dfrac{\theta }{2}%
\end{array}%
\right) .  \notag
\end{eqnarray}%
Finally, by multiplication of the matrices in (\ref{po20}) and (\ref{po21})
one gets%
\begin{equation}
\left( 1-\eta ^{2}\right) \left(
\begin{array}{cccc}
\eta ^{2}\cos ^{2}\dfrac{\theta }{2} & \frac{1}{2}\eta ^{2}\sin \theta \
e^{- \iun \varphi } & -\eta \cos ^{2}\dfrac{\theta }{2} & -\frac{1}{2}\eta \sin
\theta \ e^{- \iun \varphi } \\
\frac{1}{2}\eta ^{2}\sin \theta \ e^{ \iun \varphi } & \eta ^{2}\sin ^{2}\dfrac{%
\theta }{2} & -\frac{1}{2}\eta \sin \theta \ e^{ \iun \varphi } & -\eta \sin ^{2}%
\dfrac{\theta }{2} \\
\eta \cos ^{2}\dfrac{\theta }{2} & \frac{1}{2}\eta \sin \theta \
e^{- \iun \varphi } & -\cos ^{2}\dfrac{\theta }{2} & -\frac{1}{2}\sin \theta \
e^{- \iun \varphi } \\
\frac{1}{2}\eta \sin \theta \ e^{ \iun \varphi } & \eta \sin ^{2}\dfrac{\theta }{2%
} & -\frac{1}{2}\sin \theta \ e^{ \iun \varphi } & -\sin ^{2}\dfrac{\theta }{2}%
\end{array}%
\right) . \label{po22}
\end{equation}%
This matrix is equal to the left hand side, namely the tensor product $\left[
\rho _{-}\left( p\right) \right] _{\alpha \beta },$ of the relation (\ref%
{po14aa}) for the bi-spinor (\ref{po19a}), up to the required constant. The
case $\lambda =1/2$ can be verified in a similar fashion, say with the help
of a computer algebra system. Further details are left to the reader.

\textbf{Note}. The bi-spinors (\ref{h5})--(\ref{h6}) and (\ref{po19a})--(\ref%
{po19b}) are linearly independent because the determinant:%
\begin{equation}
\left\vert
\begin{array}{cccc}
\cos \dfrac{\theta }{2}e^{- \iun \varphi /2} & \sin \dfrac{\theta }{2}%
e^{- \iun \varphi /2} & \eta \sin \dfrac{\theta }{2}e^{- \iun \varphi /2} & \eta \cos
\dfrac{\theta }{2}e^{- \iun \varphi /2} \\
\sin \dfrac{\theta }{2}e^{ \iun \varphi /2} & -\cos \dfrac{\theta
	}{2}e^{ \iun \varphi
/2} & -\eta \cos \dfrac{\theta }{2}e^{ \iun \varphi /2} & \eta \sin \dfrac{\theta
}{2}e^{ \iun \varphi /2} \\
\eta \cos \dfrac{\theta }{2}e^{- \iun \varphi /2} & -\eta \sin \dfrac{\theta }{2}%
e^{- \iun \varphi /2} & -\sin \dfrac{\theta }{2}e^{- \iun \varphi /2} & \cos \dfrac{%
\theta }{2}e^{- \iun \varphi /2} \\
\eta \sin \dfrac{\theta }{2}e^{ \iun \varphi /2} & \eta \cos \dfrac{\theta }{2}%
e^{ \iun \varphi /2} & \cos \dfrac{\theta }{2}e^{ \iun \varphi /2} & \sin \dfrac{%
\theta }{2}e^{ \iun \varphi /2}%
\end{array}%
\right\vert =\left( 1-\eta ^{2}\right) ^{2} \label{po23}
\end{equation}%
is not zero when $\left\vert \eta \right\vert <1\/ $ (see our complementary Mathematica file).  $\qquad \blacksquare $

\textbf{Note}. The formulas (\ref{po14})--(\ref{po14aa}) permit one to make
covariant polarization calculations in terms of traces \cite{MicheWightman55} (see also Appendix~C). %
In the original form, for the corresponding projection operators, one can
write that
\begin{eqnarray}
\frac{u_{\alpha }^{\left( \lambda \right) }\left( p\right) \overline{u}%
_{\beta }^{\left( \lambda \right) }\left( p\right) }{\overline{u}^{\left(
\lambda \right) }\left( p\right) u^{\left( \lambda \right) }\left( p\right) }
&=&\frac{1}{4mc}\left( mc+\widehat{p}\right) \left( I-\gamma _{5}\widehat{a}%
\right) ,  \label{po24} \\
\frac{v_{\alpha }^{\left( \lambda \right) }\left( p\right) \overline{v}%
_{\beta }^{\left( \lambda \right) }\left( p\right) }{\overline{v}^{\left(
\lambda \right) }\left( p\right) v^{\left( \lambda \right) }\left( p\right) }
&=&\frac{1}{4mc}\left( mc-\widehat{p}\right) \left( I-\gamma _{5}\widehat{a}%
\right) ,  \label{po25}
\end{eqnarray}%
in view of (\ref{po2}) and (\ref{po8a}), for the positive and negative
energy eigenvalues, respectively. According to our calculations, in
components, it means that%
\begin{equation}
\frac{u_{\alpha }^{\left( \pm 1/2\right) }\left( p\right) \overline{u}%
_{\beta }^{\left( \pm 1/2\right) }\left( p\right) }{\overline{u}^{\left( \pm
1/2\right) }\left( p\right) u^{\left( \pm 1/2\right) }\left( p\right) }=%
\frac{1}{4mc}\sum_{\delta =1}^{4}\left( mc+\widehat{p}\right) _{\alpha
\delta }\left( I\mp \gamma _{5}\widehat{a}\right) _{\delta \beta },\qquad E>0
\label{po26}
\end{equation}%
and%
\begin{equation}
\frac{v_{\alpha }^{\left( \pm 1/2\right) }\left( p\right) \overline{v}%
_{\beta }^{\left( \pm 1/2\right) }\left( p\right) }{\overline{v}^{\left( \pm
1/2\right) }\left( p\right) v^{\left( \pm 1/2\right) }\left( p\right) }=%
\frac{1}{4mc}\sum_{\delta =1}^{4}\left( mc-\widehat{p}\right) _{\alpha
\delta }\left( I\mp \gamma _{5}\widehat{a}\right) _{\delta \beta },\qquad E<0
\label{po27}
\end{equation}%
(all four relations have been verified by direct matrix multiplications with the aid of the Mathematica computer algebra system).
\qquad $\blacksquare $

\subsection{A Covariant Approach}

Let us consider, for example, the case of positive energy eigenvalues $E>0$
and the positive polarization $\lambda =1/2.$ In view of (\ref{h3}), one gets%
\begin{equation}
mc+\widehat{p}=\frac{2mc}{1-\eta ^{2}}\left(
\begin{array}{cc}
\mathbf{1} & -\eta \left( \boldsymbol{\sigma \cdot n}\right) \\
\eta \left( \boldsymbol{\sigma \cdot n}\right) & -\eta ^{2}\mathbf{1}%
\end{array}%
\right) .  \label{ca1}
\end{equation}%
When $\boldsymbol{p}=\left\vert \boldsymbol{p}\right\vert \ \boldsymbol{n},$
equations (\ref{rd10}) take the form%
\begin{eqnarray}
a^{0} &=&\frac{\left\vert \boldsymbol{p}\right\vert }{mc}=\frac{2\eta }{%
1-\eta ^{2}},  \label{ca2} \\
\boldsymbol{a} &=&\boldsymbol{n}+\frac{\boldsymbol{n\left\vert \boldsymbol{p}%
\right\vert }^{2}}{m\left( E+mc^{2}\right) }=\frac{1+\eta ^{2}}{1-\eta ^{2}}%
\boldsymbol{n}.  \label{ca3}
\end{eqnarray}%
Therefore,%
\begin{equation}
\boldsymbol{\sigma \cdot a}=\frac{1+\eta ^{2}}{1-\eta ^{2}}\left(
\boldsymbol{\sigma \cdot n}\right) .  \label{ca4}
\end{equation}%
By (\ref{po15}) we obtain:%
\begin{equation}
I-\gamma _{5}\widehat{a}=\frac{1}{1-\eta ^{2}}\left(
\begin{array}{cc}
\left( 1-\eta ^{2}\right) \mathbf{1}+\left( 1+\eta ^{2}\right) \left(
\boldsymbol{\sigma \cdot n}\right) & -2\eta \mathbf{1} \\
2\eta \mathbf{1} & \left( 1-\eta ^{2}\right) \mathbf{1}-\left( 1+\eta
^{2}\right) \left( \boldsymbol{\sigma \cdot n}\right)%
\end{array}%
\right) ,  \label{ca5}
\end{equation}%
which is equivalent to our previous result (\ref{po17}) in view of (\ref{s6}%
).

Now, the product is given as follows%
\begin{eqnarray}
&&\left( mc+\widehat{p}\right) \left( I-\gamma _{5}\widehat{a}\right) =\frac{%
2mc}{\left( 1-\eta ^{2}\right) ^{2}}\left(
\begin{array}{cc}
\mathbf{1} & -\eta \left( \boldsymbol{\sigma \cdot n}\right)  \\
\eta \left( \boldsymbol{\sigma \cdot n}\right)  & -\eta ^{2}\mathbf{1}%
\end{array}%
\right)   \label{ca6} \\
&&\times \left(
\begin{array}{cc}
\left( 1-\eta ^{2}\right) \mathbf{1}+\left( 1+\eta ^{2}\right) \left(
\boldsymbol{\sigma \cdot n}\right)  & -2\eta \mathbf{1} \\
2\eta \mathbf{1} & \left( 1-\eta ^{2}\right) \mathbf{1}-\left( 1+\eta
^{2}\right) \left( \boldsymbol{\sigma \cdot n}\right)
\end{array}%
\right) .  \notag
\end{eqnarray}%
As a result of the matrix multiplication, perfored in a $2\times 2$ block
form, we finally obtain with the help of (\ref{s8ab}) that%
\begin{eqnarray*}
&&\left(
\begin{array}{cc}
\mathbf{1} & -\eta \left( \boldsymbol{\sigma \cdot n}\right)  \\
\eta \left( \boldsymbol{\sigma \cdot n}\right)  & -\eta ^{2}\mathbf{1}%
\end{array}%
\right) \left(
\begin{array}{cc}
\left( 1-\eta ^{2}\right) \mathbf{1}+\left( 1+\eta ^{2}\right) \left(
\boldsymbol{\sigma \cdot n}\right)  & -2\eta \mathbf{1} \\
2\eta \mathbf{1} & \left( 1-\eta ^{2}\right) \mathbf{1}-\left( 1+\eta
^{2}\right) \left( \boldsymbol{\sigma \cdot n}\right)
\end{array}%
\right)  \\
&=&\left( 1-\eta ^{2}\right) \left(
\begin{array}{cc}
\mathbf{1}+\boldsymbol{\sigma \cdot n} & -\eta \left( \mathbf{1}+\boldsymbol{%
\sigma \cdot n}\right)  \\
\eta \left( \mathbf{1}+\boldsymbol{\sigma \cdot n}\right)  & -\eta
^{2}\left( \mathbf{1}+\boldsymbol{\sigma \cdot n}\right)
\end{array}%
\right) =2\left( 1-\eta ^{2}\right) \left(
\begin{array}{cc}
\rho ^{\left( 1/2\right) }\left( \boldsymbol{\boldsymbol{n}}\right)  & -\eta
\rho ^{\left( 1/2\right) }\left( \boldsymbol{\boldsymbol{n}}\right)  \\
\eta \rho ^{\left( 1/2\right) }\left( \boldsymbol{\boldsymbol{n}}\right)  &
-\eta ^{2}\rho ^{\left( 1/2\right) }\left( \boldsymbol{\boldsymbol{n}}%
\right)
\end{array}%
\right)
\end{eqnarray*}%
provided $\left( \boldsymbol{\sigma \cdot n}\right) ^{2}=\mathbf{1}.$ This
is equivalent to the first relation (\ref{po24}).
In covariant form, one gets%
\begin{equation}
\left( mc+\widehat{p}\right) \left( I-\gamma _{5}\widehat{a}\right)
=mc\left( I-\gamma _{5}\widehat{a}\right) +\widehat{p}-\widehat{p}\gamma _{5}%
\widehat{a}.  \label{ca7}
\end{equation}%
Here,%
\begin{equation}
\widehat{p}\gamma _{5}\widehat{a}=p_{\mu }a_{\nu }\left( \gamma ^{\mu
}\gamma _{5}\gamma ^{\nu }\right) =-p_{\mu }a_{\nu }\gamma _{5}\left( \gamma
^{\mu }\gamma ^{\nu }\right) ,  \label{ca8}
\end{equation}%
in view of the anticomutator property, $\gamma _{5}\gamma ^{\nu }+\gamma
^{\mu }\gamma _{5}=0.$ Using the substitution,%
\begin{eqnarray}
\gamma ^{\mu }\gamma ^{\nu } &=&\frac{1}{2}\left( \gamma ^{\mu }\gamma ^{\nu
}-\gamma ^{\nu }\gamma ^{\mu }\right) +\frac{1}{2}\left( \gamma ^{\mu
}\gamma ^{\nu }+\gamma ^{\nu }\gamma ^{\mu }\right)   \notag \\
&=&\frac{1}{2}\left( \gamma ^{\mu }\gamma ^{\nu }-\gamma ^{\nu }\gamma ^{\mu
}\right) +g^{\mu \nu }  \label{ca9}
\end{eqnarray}%
by (\ref{rd3a}), we obtain that%
\begin{equation}
\widehat{p}\gamma _{5}\widehat{a}=-\Sigma ^{\mu \nu }p_{\mu }a_{\nu }-p_{\mu
}a^{\mu },  \label{ca10}
\end{equation}%
where $p_{\mu }a^{\mu }=0$ in view of (\ref{rd12a}) and by the definition
\begin{equation}
\Sigma ^{\mu \nu }=\frac{1}{2}\left( \gamma ^{\mu }\gamma ^{\nu }-\gamma
^{\nu }\gamma ^{\mu }\right) =\left(
\begin{array}{cccc}
0 & \alpha _{1} & \alpha _{2} & \alpha _{3} \\
-\alpha _{1} & 0 & - \iun \Sigma _{3} &  \iun \Sigma _{2} \\
-\alpha _{2} &  \iun \Sigma _{3} & 0 & - \iun \Sigma _{1} \\
-\alpha _{3} & - \iun \Sigma _{2} &  \iun \Sigma _{1} & 0%
\end{array}%
\right) ,\quad \boldsymbol{\Sigma }=\left(
\begin{array}{cc}
\boldsymbol{\sigma } & \mathbf{0} \\
\mathbf{0} & \boldsymbol{\sigma }%
\end{array}%
\right) \label{ca10a}
\end{equation}%
As a result, one can write%
\begin{equation}
\widehat{p}\gamma _{5}\widehat{a}=-\gamma _{5}\left( \widehat{pa}\right)
,\qquad \widehat{pa}:=\Sigma ^{\mu \nu }p_{\mu }a_{\nu }=-p_{0}\left(
\boldsymbol{\alpha \cdot a}\right) +a_{0}\left( \boldsymbol{\alpha \cdot p}%
\right) - \iun \boldsymbol{\Sigma \cdot }\left( \boldsymbol{p}\times \boldsymbol{a%
}\right)   \label{ca11}
\end{equation}%
(see, for example, \cite{BerLifPit}, \cite{Kr:Lan:Sus16}, or \cite{Moskalev}%
). Our relation (\ref{ca7}) take the form%
\begin{equation}
\left( mc+\widehat{p}\right) \left( I-\gamma _{5}\widehat{a}\right)
=mc\left( I-\gamma _{5}\widehat{a}\right) +\widehat{p}+\gamma _{5}\widehat{pa%
},  \label{ca12}
\end{equation}%
where%
\begin{equation}
\widehat{p}=\frac{mc}{1-\eta ^{2}}\left(
\begin{array}{cc}
\left( 1+\eta ^{2}\right) \mathbf{1} & -2\eta \left( \boldsymbol{\sigma
\cdot n}\right)  \\
2\eta \left( \boldsymbol{\sigma \cdot n}\right)  & -\left( 1+\eta
^{2}\right) \mathbf{1}%
\end{array}%
\right) ,\qquad \gamma _{5}\widehat{pa}=mc\left(
\begin{array}{cc}
\boldsymbol{\sigma \cdot n}
 & \mathbf{0} \\
\mathbf{0} &
\boldsymbol{\sigma \cdot n}
\end{array}%
\right) .
\end{equation}%
Using (\ref{ca5}), one can reduce the last but one relation to the required
form by an elementary calculation. All of the remaining cases can be
verified in a similar fashion. We leave the details to the reader. In
summary, we have re-calculated the projection operators (\ref{po26})--(\ref%
{po27}) in $4\times 4$ and in $2\times 2$ block matrix forms, as well as in
the covariant form in detail.

\section{The Charge Conjugation\/}

According to our explicit forms of bi-spinors (\ref{h5})--(\ref{h6}) and (\ref%
{po19a})--(\ref{po19b}), one gets%
\begin{eqnarray}
 \iun \gamma ^{2}\left( v_{+}^{\left( 1/2\right) }\right) ^{\ast }
&=&v_{-}^{\left( 1/2\right) },  \label{cc1} \\
 \iun \gamma ^{2}\left( v_{+}^{\left( -1/2\right) }\right) ^{\ast }
&=&v_{-}^{\left( -1/2\right) }.  \label{cc2}
\end{eqnarray}%
This fact implies the well-known property of charge conjugation for free
Dirac particles:%
\begin{equation}
\mathbf{C}\left[ \psi _{+}\left( x\right) \right] ^{\ast }=\psi _{-}\left(
x\right) ,\qquad  \iun \gamma ^{2}\left[ u^{\left( \lambda \right) }\left(
p\right) \right] ^{\ast }=v^{\left( \lambda \right) }\left( p\right)
\label{cc3}
\end{equation}%
provided%
\begin{equation}
\psi _{+}\left( x\right) =u^{\left( \lambda \right) }\left( p\right)
e^{- \iun \left( p_{\mu }x^{\mu }\right) /\hbar },\qquad \psi _{-}\left( x\right)
=v^{\left( \lambda \right) }\left( p\right) e^{ \iun \left( p_{\mu }x^{\mu
}\right) /\hbar }.  \label{cc4}
\end{equation}%
(More details can be found in \cite{AkhBer}, \cite{BerLifPit}, \cite%
{Moskalev}.)

\newpage

\section{Appendix A: Some Required Facts From Matrix Algebra\/}

When two partitioned matrices have the same shape and their diagonal blocks
are square matrices of equal size, then the following multiplication rule
holds:%
\begin{equation}
\left(
\begin{array}{cc}
A_{1} & B_{1} \\
C_{1} & D_{1}%
\end{array}%
\right) \left(
\begin{array}{cc}
A_{2} & B_{2} \\
C_{2} & D_{2}%
\end{array}%
\right) =\left(
\begin{array}{cc}
A_{1}A_{2}+B_{1}C_{2} & A_{1}B_{2}+B_{1}D_{2} \\
C_{1}A_{2}+D_{1}C_{2} & C_{1}B_{2}+D_{1}D_{2}%
\end{array}%
\right) \label{A1}
\end{equation}%
that is similar to multiplication of $2\times 2$ matrices \cite%
{GantmacherMat}.

Let us consider a determinant partitioned into four blocks:%
\begin{equation}
\Delta =\left\vert
\begin{array}{cc}
A & B \\
C & D%
\end{array}%
\right\vert \label{A2}
\end{equation}%
where $A$ and $D$ are square matrices. There are formulas of Schur, which
reduce the computation of a determinant of order $2n$ to the computation of
a determinant of order $n$:%
\begin{equation}
\Delta =\left\vert AD-CB\right\vert \label{A3a}
\end{equation}%
provided $AC=CA$ or%
\begin{equation}
\Delta =\left\vert AD-BC\right\vert \label{A3b}
\end{equation}%
provided $CD=DC$ (see \cite{GantmacherMat}, pp.~45--46).

Finally, the following result is also important.

\begin{theorem}
If a rectangular matrix $R$ is represented in partitioned form%
\begin{equation}
R=\left(
\begin{array}{cc}
A & B \\
C & D%
\end{array}%
\right) , \label{A4}
\end{equation}%
where $A$ is a square non-singular matrix of order $n$ ($\left\vert
A\right\vert \neq 0$), then the rank of $R$ is equal to $n$ if and only if
\begin{equation}
D=CA^{-1}B. \label{A5}
\end{equation}
\end{theorem}

(See proof of Theorem~4 on p.~47 in \cite{GantmacherMat}.)

\newpage

\section{Appendix B: Fermi's Lecture Notes on Relativistic Free Electron\/}

\setcounter{section}{34}


Here we present both Enrico Fermi's original lecture notes for Physics 341/342: Quantum Mechanics at the University of Chicago \cite{Fermi} in the
winter and spring of 1954  -- two quarters and approximately sixty lectures -- along with a typeset and very lightly edited
transcription. 
At the end of each lecture, Fermi would always make up a problem, which was usually closely related to what he had just talked
about that day. For further details on some subjects, Fermi occasionally mentioned Leonard Schiff's book, 
{\it{Quantum Mechanics\/}}, First Edition \cite{Schiff}.

Note that we have preserved Fermi's notation conventions. For
example, where a typeset manuscript would typically use boldface
for a vector, Fermi's handwritten notes naturally use a
superscripted arrow; here, we maintain the arrow and use singular
vertical lines for vectors and matrices.

At the very beginning of his distinguished career, Fermi held a temporary job for the year 1923-24 at the University of Rome,
teaching a course of mathematics for chemists and biologists.
From 1924 to 1926, Fermi lectured in mathematical physics and mechanics at the University of Florence. 
At that time, he had mastered Schr{\"o}dinger's theory from the original publications and explained them to his students in private seminars;
later he recast some of Dirac's papers in more familiar form, in part for didactical reasons \cite{Fermi}.
Then Fermi became the first professor of theoretical physics
at the University of Rome, where he taught for 12 years since 1926 \cite{Bruzz}, \cite{Segre}.

All his life Fermi kept passion for teaching.
He gave numerous courses at the University of Rome, at Columbia University, and at the University of Chicago.
After his last quantum mechanics course at the University of Chicago in the winter and spring semesters
and before his untimely death in November, during the summer of 1954 Fermi went to Europe. He had prepared a course on pions and nucleons,
which he delivered at the Villa Monastero (in Varenna on Lake Como) at the summer school of the Italian Physical Society,
which is now named after him. He also visited the French summer school at Les Houches near Chamonix, and lectured there \cite{Segre}.
\footnote{
From Britannica:
Enrico Fermi, (born Sept. 29, 1901, Rome, Italy -- died Nov. 28, 1954, Chicago, Illinois, U.~S.~A.), Italian-born American scientist who was one of the chief architects of the nuclear age. He developed the mathematical statistics required to clarify a large class of subatomic phenomena, explored nuclear transformations caused by neutrons, and directed the first controlled chain reaction involving nuclear fission. He was awarded the 1938 Nobel Prize for Physics, and the Enrico Fermi Award of the U.S. Department of Energy is given in his honour. Fermilab, the National Accelerator Laboratory, in Illinois, is named for him, as is fermium, element number 100.
{\url{https://www.britannica.com/biography/Enrico-Fermi}}}

\newpage

\begin{figure}[h!]
	\includegraphics[width=0.75 \linewidth]{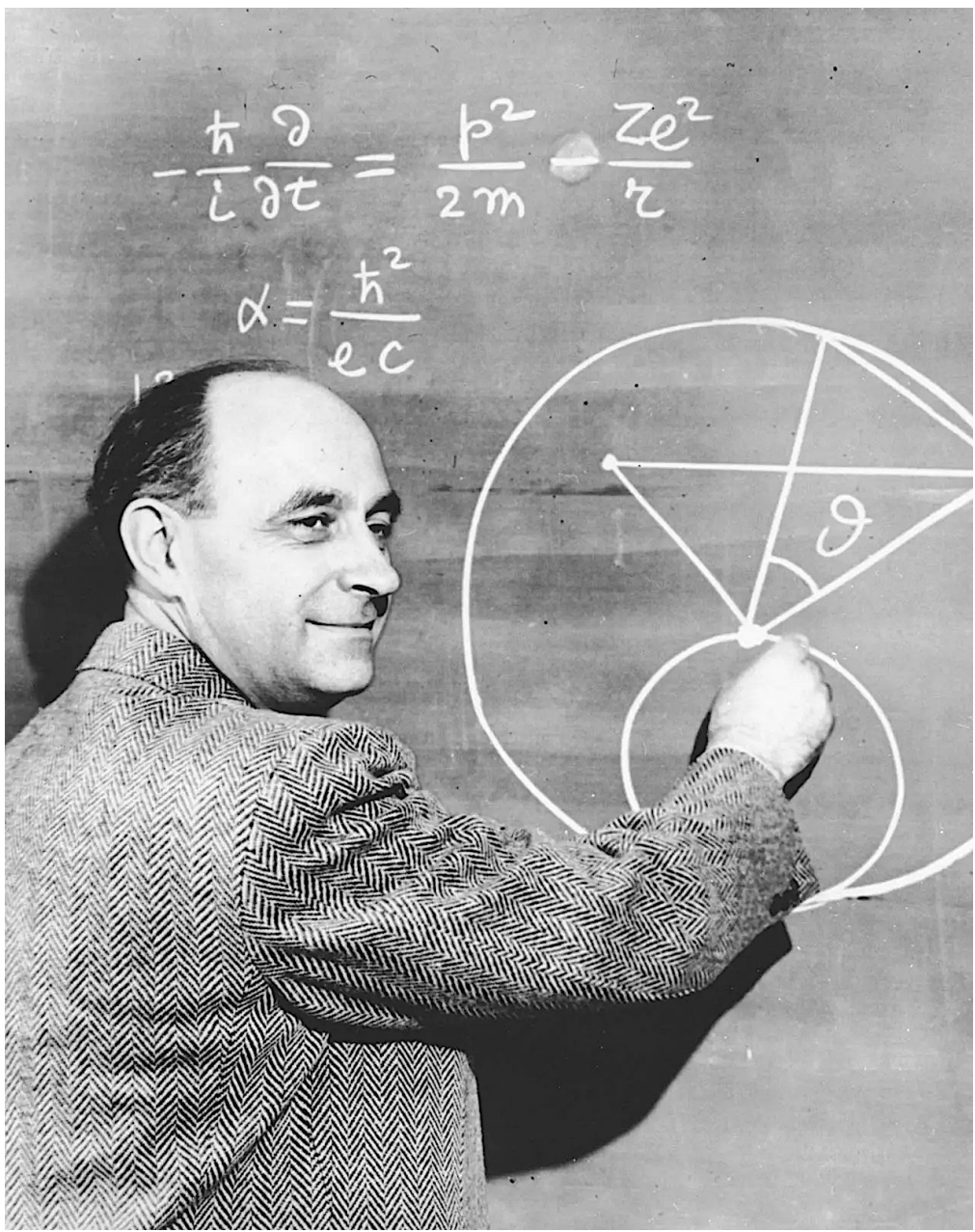}
	\caption{Enrico Fermi teaching quantum mechanics.
	From Encyclopedia of Britannica}
\end{figure}

\vfill
\eject
\newpage

\includepdf[pagecommand={}, height=\textheight, keepaspectratio, pages={1}]{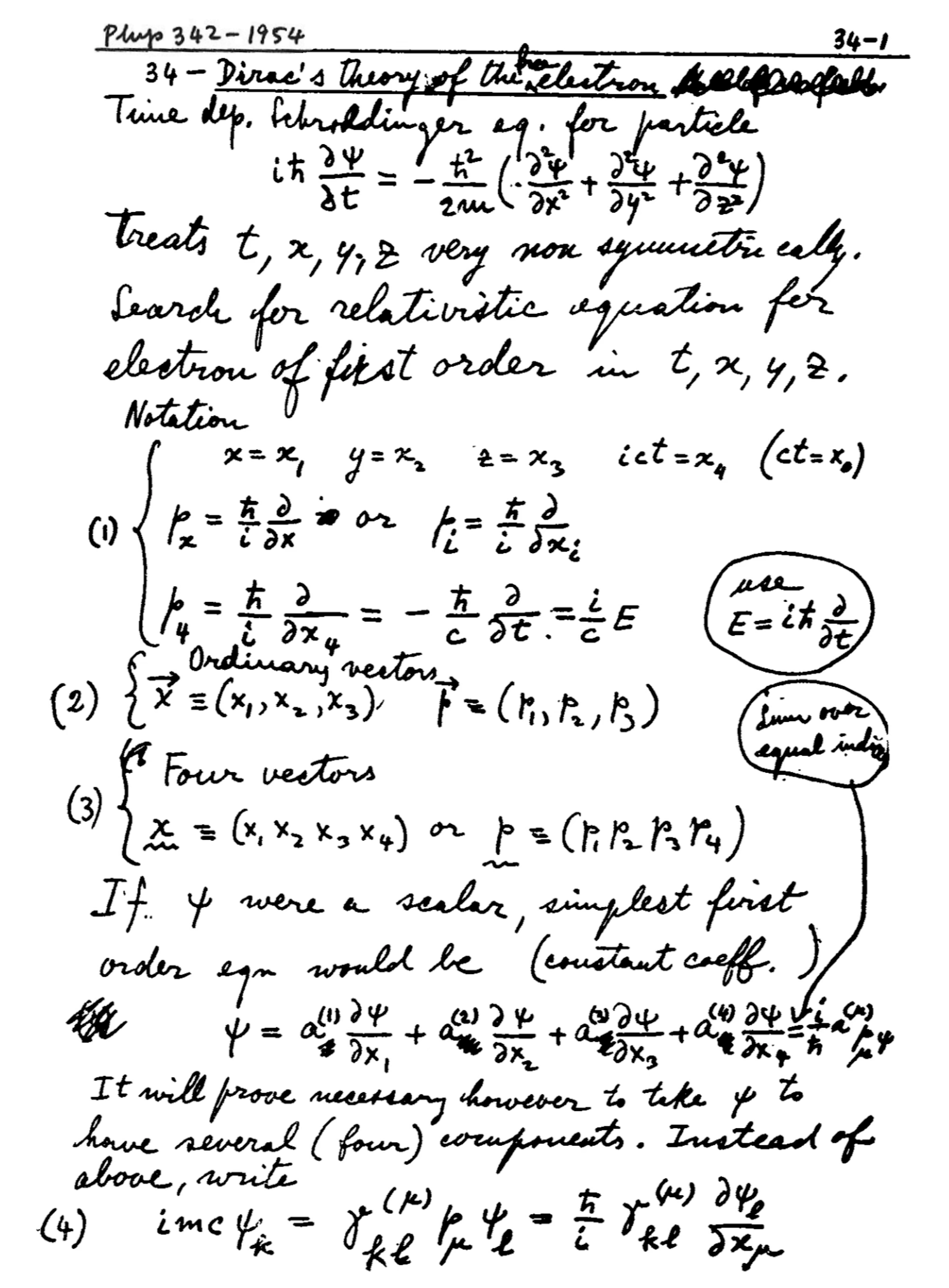}

\textbf{34 -- Dirac's theory of the free electron\/}

\noindent
The time-dependent Schr{\"o}dinger equation for a particle
\begin{equation*}
	 \iun  \hslash \diffp{\psi}{t}
	=
	- \frac{\hslash}{2m}
	\left(
		\diffp[2]{\psi}{x}
		+
		\diffp[2]{\psi}{y}
		+
		\diffp[2]{\psi}{z}
	\right)
\end{equation*}
treats $t, x, y, z$ non-symmetrically.

We will search for the relativistic equation for an electron of
first order in $t, x, y, z$.

Notation:

\begin{equation}
	\begin{cases}
		x = x_1, \, y = x_2, \, z = x_3, \, {\iun} ct = x_4 \; (ct=x_0) \\
		p_x = \frac{\hslash}{\iun} \diffp{}{x} \; \operatorname{or} \; p_i = \frac{\hslash}{\iun} \diffp{}{{x_{i}}} \; \; \\
		p_4 = \frac{\hslash}{\iun} \diffp{}{{x_{4}}} = - \frac{\hslash}{c}\diffp{}{t} = \frac{\iun}{c}E \; \; \; \; \left(\operatorname{use} E = \iun\hslash\diffp{}{t} \right)
	\end{cases}
\end{equation}

Ordinary vectors:

\begin{equation}
	\begin{cases}
		\vec{x}
		\equiv
		\begin{bmatrix} x_1 & x_2 & x_3 \end{bmatrix}
		\, \operatorname{or} \,
		\vec{p}
		\equiv
		\begin{bmatrix} p_1 & p_2 & p_3 \end{bmatrix}
	\end{cases}
\end{equation}

Four vectors:

\begin{equation}
	\begin{cases}
		\fourvec{x}
		\equiv
		\begin{bmatrix} x_1 & x_2 & x_3 & x_4 \end{bmatrix}
		\, \operatorname{or} \,
		\fourvec{p}
		\equiv
		\begin{bmatrix} p_1 & p_2 & p_3 & p_4 \end{bmatrix}
	\end{cases}
\end{equation}

If $\psi$ were a scalar, the simplest first order equation with
constant coefficients would be
\begin{equation*}
	\psi =
		a^{(1)} \diffp{\psi}{{x_{1}}} +
		a^{(2)} \diffp{\psi}{{x_{2}}} +
		a^{(3)} \diffp{\psi}{{x_{3}}} +
		a^{(4)} \diffp{\psi}{{x_{4}}}
		=
		\frac{\iun}{\hslash}a^{(\mu)}p_{\mu}\psi,
\end{equation*}
where we sum over equal indices.

\noindent
However, it will prove necessary to take $\psi$ to have four
components. Instead of the above equation, we write:
\begin{equation}
	\iun mc\psi_k
	=
	\gamma^{(\mu)}_{k\ell} p_\mu \psi_\ell
	=
	\frac{\hslash}{\iun} \gamma^{(\mu)}_{k\ell}
	\diffp{{\psi_{\ell}}}{{x_{\mu}}}
\end{equation}

\includepdf[pagecommand={}, height=\textheight, keepaspectratio, pages={2}]{Fermi.pdf}

\noindent
In matrix notation, we have $\psi$ as a vertical slot of four
elements and $\gamma_\mu = \begin{Vmatrix} \gamma^{(\mu)}_{k\ell}
\end{Vmatrix}$, which is a square four-by-four matrix:
\begin{align}
	\iun mc\psi &= \gamma_\mu p_\mu \psi \; (\operatorname{sums\;over} \mu) \nonumber \\
	&= \frac{\hslash}{\iun} \gamma_\mu \diffp{\psi}{{x_{\mu}}}
\end{align}
where $p_\mu = \dfrac{\hslash}{\iun} \diffp{}{{x_{\mu}}}$ operates on
the dependence of $\psi$ on $\fourvec{x}$ and $\gamma_\mu$
operates on an internal variable similar to the spin variable of
Pauli --- however, with four components, as will be seen. It
therefore follows that
\begin{equation}
	\begin{cases}
		\gamma_\mu \operatorname{commutes\;with} p_\nu \operatorname{and \;}
		x_\nu.
	\end{cases}
\end{equation}

\noindent
From (5), we have
\begin{equation*}
	(\iun mc)^2\psi = (\gamma_\mu p_\mu)^2 \psi,
\end{equation*}
or (omitting $\psi$)
\begin{eqnarray*}
	-m^2c^2 &=&
	\gamma_1^2 p_1^2 +
	\gamma_2^2 p_2^2 +
	\gamma_3^2 p_3^2
	- \gamma_4^2 \frac{E^2}{c^2} \\
	&+& (\gamma_1\gamma_2 + \gamma_2\gamma_1)p_1 p_2
	+ \operatorname{\; similar\;terms},
\end{eqnarray*}
where we use from (1) that $p_4^2 = -\dfrac{E^2}{c^2}$.

\noindent
This can be identified with the relativistic momentum energy
relation
\begin{equation}
	m^2c^2 + \vec{p}^{\;2} = \frac{E^2}{c^2}
\end{equation}
by postulating
\begin{equation}
	\gamma_1^2 = \gamma_2^2 = \gamma_3^2 = \gamma_4^2 = 1
	\;\;\;\;\;\;\;\;\;
	\gamma_\mu\gamma_\nu + \gamma_\nu\gamma_\mu = 0
	\; \operatorname{for} \; \mu \neq \nu.
\end{equation}

\includepdf[pagecommand={}, height=\textheight, keepaspectratio, pages={3}]{Fermi.pdf}

\noindent
One finds that the lowest order matrices for which (8) can be
fulfilled is the fourth. For order four, there are many solutions
that are essentially equivalent. We choose the standard solution:
\begin{equation}
	\gamma_1 =
	\begin{vmatrix}
		0 & 0 & 0 & -\iun \\
		0 & 0 & -\iun & 0 \\
		0 & \iun & 0 & 0 \\
		\iun & 0 & 0 & 0 \\
	\end{vmatrix}
	;\;
	\gamma_2 =
	\begin{vmatrix}
		0 & 0 & 0 & -1 \\
		0 & 0 & 1 & 0 \\
		0 & 1 & 0 & 0 \\
		-1 & 0 & 0 & 0 \\
	\end{vmatrix}
	;\;
	\gamma_3 =
	\begin{vmatrix}
		0 & 0 & -\iun & 0 \\
		0 & 0 & 0 & \iun \\
		\iun & 0 & 0 & 0 \\
		0 & -\iun & 0 & 0 \\
	\end{vmatrix}
\end{equation}
and
\begin{equation}
	\beta = \gamma_4 =
	\begin{vmatrix}
		1 & 0 & 0 & 0 \\
		0 & 1 & 0 & 0 \\
		0 & 0 & -1 & 0 \\
		0 & 0 & 0 & -1 \\
	\end{vmatrix}.
\end{equation}

Note that $\gamma_1, \gamma_2,$ and $\gamma_3$ act in many ways as
the components of a vector; therefore, they will be denoted by
\begin{equation}
	\vec{\gamma} \equiv (\gamma_1, \gamma_2, \gamma_3)
	\; \operatorname{also} \;
	\fourvec{\gamma} \equiv (\gamma_1, \gamma_2, \gamma_3,
	\gamma_4).
\end{equation}

Now, (5) becomes
\begin{equation}
	\iun mc\psi = (\vec{\gamma} \cdot \vec{p} + \frac{{\iun}}{c} E\gamma_4)
	\psi = \fourvec{\gamma} \cdot \fourvec{p} \, \psi.
\end{equation}

We now multiply to left by $\gamma_4 = \beta$ using
$\gamma_4^2 = \beta^2 = 1$ to get
\begin{equation}
	\boxed{
		E\psi = (mc^2\beta + c\vec{\alpha} \cdot \vec{p})\psi
	}
\end{equation}
where
\begin{equation}
	\vec{\alpha} = \iun \beta \vec{\gamma} \quad
	(\operatorname{or} \alpha_1 = \iun\beta\gamma_1, \, \alpha_2 =
	\iun\beta\gamma_2, \, \alpha_3=\iun\beta\gamma_3)
\end{equation}
and
\begin{equation}
	\alpha_1 =
	\begin{vmatrix}
		0 & 0 & 0 & 1 \\
		0 & 0 & 1 & 0 \\
		0 & 1 & 0 & 0 \\
		1 & 0 & 0 & 0
	\end{vmatrix}
	;\;
	\alpha_2 =
	\begin{vmatrix}
		0 & 0 & 0 & -\iun \\
		0 & 0 & \iun & 0 \\
		0 & -\iun & 0 & 0 \\
		\iun & 0 & 0 & 0
	\end{vmatrix}
	;\;
	\alpha_3 =
	\begin{vmatrix}
		0 & 0 & 1 & 0 \\
		0 & 0 & 0 & -1 \\
		-1 & 0 & 0 & 0 \\
		0 & -1 & 0 & 0
	\end{vmatrix}.
\end{equation}

\includepdf[pagecommand={}, height=\textheight, keepaspectratio, pages={4}]{Fermi.pdf}

This equation has the following properties, which should be
checked directly:
\begin{equation}
	\beta^2 = \alpha_1^2 = \alpha_2^2 = \alpha_3^2 =1
\end{equation}
\begin{equation}
	\begin{cases}
		\beta\alpha_1 + \alpha_1\beta = 0
		\;\;\;\;
		\beta\alpha_2 + \alpha_2\beta = 0
		\;\;\;\;
		\beta\alpha_3 + \alpha_3\beta = 0
		\\
		\alpha_1\alpha_2 + \alpha_2\alpha_1 = 0
		\;\;\;\;
		\alpha_2\alpha_3 + \alpha_3\alpha_2 = 0
		\;\;\;\;
		\alpha_3\alpha_1 + \alpha_1\alpha_3 = 0
	\end{cases}
\end{equation}
\begin{equation}
	\begin{cases}
		\beta + \textrm{the } \alpha^{,}\textrm{s have square} = \textrm{unit matrix;} \\
		\beta + \textrm{the } \alpha^{,}\textrm{s anticommute with each other; and} \\
		\beta + \textrm{the } \alpha^{,}\textrm{s are Hermitian.}
	\end{cases}
\end{equation}

\noindent
One can prove that all the physical consequences of (13) do not
depend on the special choices of (10) and (15) of $\alpha_1,
\alpha_2, \alpha_3,$ and $\beta$. They would be the same if a
different set of four $4\times4$ matrices with the specifications
of (18) had been chosen. In particular, it is possible by unitary
transformation to interchange the roles of the four matrices;
therefore, their differences are merely apparent.

\begin{equation}
	\begin{cases}
		\textrm{Check that for each of the seven matrices} \\
		\gamma_4 = \beta, \, \alpha_1, \, \alpha_2, \, \alpha_3, \, \gamma_1, \, \gamma_2, \, \gamma_3 \\
		\textrm{the eigenvalues are +1  twice and -1  twice.}
	\end{cases}
\end{equation}

\includepdf[pagecommand={}, height=\textheight, keepaspectratio, pages={5}]{Fermi.pdf}

Observe that (13) may also be written as
\begin{equation}
	E\psi = H\psi
\end{equation}
where
\begin{equation}
	\begin{cases}
		H = \operatorname{Hamiltonian} \\
		H = mc^2\beta + c\vec{\alpha} \cdot \vec{p}
	\end{cases}
\end{equation}
for $\psi = \begin{vmatrix} \psi_1 \\ \psi_2 \\ \psi_3 \\ \psi_4
\end{vmatrix}$.

As a time-independent equation, we have
\begin{equation}
	\begin{cases}
		E\psi_1 =
			mc^2 \psi_1 +
			\frac{c\hslash}{\iun}
			\left\{
				 \diffp{{\psi_{4}}}{x} -
				\iun\diffp{{\psi_{4}}}{y} +
				 \diffp{{\psi_{3}}}{z}
			\right\} \\
		E\psi_2 =
			mc^2 \psi_2 +
			\frac{c\hslash}{\iun}
			\left\{
				 \diffp{{\psi_{3}}}{x} +
				\iun\diffp{{\psi_{3}}}{y} -
				 \diffp{{\psi_{4}}}{z}
			\right\} \\
		E\psi_3 =
			mc^2 \psi_2 +
			\frac{c\hslash}{\iun}
			\left\{
				 \diffp{{\psi_{2}}}{x} -
				\iun\diffp{{\psi_{2}}}{y} +
				 \diffp{{\psi_{1}}}{z}
			\right\} \\
		E\psi_4 =
			mc^2 \psi_4 +
			\frac{c\hslash}{\iun}
			\left\{
				 \diffp{{\psi_{1}}}{x} +
				\iun\diffp{{\psi_{1}}}{y} -
				 \diffp{{\psi_{2}}}{z}
			\right\}. \\
	\end{cases}
\end{equation}

We also obtain the time-dependent Schr{\"o}dinger-type equation by $E
\to \iun\hslash \diffp{}{t}$.

For the plane wave solution, take
\begin{equation}
	\psi = \begin{vmatrix}u_1\\u_2\\u_3\\u_4\end{vmatrix}
		e^{\dfrac{\iun}{\hslash}\vec{p}\cdot\vec{x}},
\end{equation}
where $\vec{p}$ is now a numerical vector and $u_1, \  u_2,
\ u_3,$ and $u_4$ are constants.

Now, divide (22) by common exponential factors to obtain
\begin{equation}
	\begin{cases}
		E u_1 = mc^2 u_1 + c(p_x - \iun p_y) u_4 + cp_z u_3 \\
		E u_2 = mc^2 u_2 + c(p_x + \iun p_y) u_3 - cp_z u_4 \\
		E u_3 = mc^2 u_3 + c(p_x - \iun p_y) u_2 + cp_z u_1 \\
		E u_4 = mc^2 u_4 + c(p_x + \iun p_y) u_1 - cp_z u_2, \\
	\end{cases}
\end{equation}
which is a system of four homogenous linear equations with
unknowns $u_1, u_2, u_3,$ and $u_4$.

If we require a zero determinant, one finds eigenvalues of E to be
\begin{equation}
	E = +\sqrt{\mu^2 c^4 + c^2 p^2} \, \, \operatorname{twice} \,
	\operatorname{and} \;
	E = -\sqrt{\mu^2 c^4 + c^2 p^2} \, \operatorname{twice}.
\end{equation}

\includepdf[pagecommand={}, height=\textheight, keepaspectratio, pages={6}]{Fermi.pdf}

For each $\vec{p}, E$ has twice the value, $E = \sqrt{m^2 c^4 +
c^2 p^2}$, but also twice the negative value $E = -\sqrt{m^2 c^4
+ c^2 p^2}$. (Comments will follow.)

A set of four orthogonal normalized spinors $u$ is
\footnote{%
See our section 4 for a correction.}
:
\begin{equation}
	\begin{cases}
		\operatorname{For}
		E = + \sqrt{m^2 c^4 + c^2 p^2} = R \\
		u^{(1)} = \sqrt{\dfrac{mc^2+R}{2R}}
		\begin{vmatrix}
			1 \\
			0 \\
			\dfrac{cp_z}{mc^2+R} \\
			\dfrac{c(p_x + \iun p_y)}{mc^2+R}
		\end{vmatrix}
		\, \operatorname{or} \,
		u^{(2)} = \sqrt{\dfrac{mc^2+R}{2R}}
		\begin{vmatrix}
			0 \\
			1 \\
			\dfrac{c(p_x - \iun p_y)}{mc^2+R} \\
			\dfrac{-cp_z}{mc^2+R}
		\end{vmatrix}
	\end{cases}
\end{equation}
\begin{equation}
	\begin{cases}
		\operatorname{For}
		E = -R = - \sqrt{m^2 c^4 + c^2 p^2} \\
		u^{(3)} = \sqrt{\dfrac{R - mc^2}{2R}}
		\begin{vmatrix}
			\dfrac{cp_z}{R-mc^2} \\
			\dfrac{c(p_x + \iun p_y)}{R-mc^2} \\
			1 \\
			0
		\end{vmatrix}
		\, \operatorname{or} \,
		u^{(4)} = \sqrt{\dfrac{R-mc^2}{2R}}
		\begin{vmatrix}
			\dfrac{c(p_x - \iun p_y)}{R-mc^2} \\
			\dfrac{-cp_z}{R-mc^2} \\
			0 \\
			1
		\end{vmatrix}
	\end{cases}
\end{equation}

\noindent
Observe: for $\lvert p \rvert \ll m c$, the third and fourth
components of the positive energy solutions $u^{(1)}$ and
$u^{(2)}$ are very small; and the first and second components of
the negative energy solutions $u^{(3)}$ and $u^{(4)}$ are also
very small (on the order of $p/{mc})$.

\includepdf[pagecommand={}, height=\textheight, keepaspectratio, pages={7}]{Fermi.pdf}

\subsection*{Meaning of negative and positive energy levels}

\subsubsection*{The Dirac Sea; the Vacuum State; Positrons as Holes}

[See Dirac's original notes for a diagram of positive and negative
energy states.]

\noindent \begin{minipage}[c]{.175\textwidth}
  \centering
 \includegraphics[ width =0.75\textwidth]{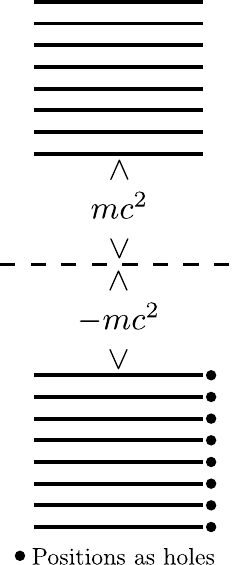}
         \end{minipage}%
        \hfil
\begin{minipage}[c]{.825\textwidth}
Momentum and energy of the positron are $(-\vec{p} + -\vec{E})$ of
the ``hole'' state.

Electron states (spin up and spin down) are represented by

\begin{equation}
	\begin{cases}
		u^{(1)} e^{\frac{\iun }{\hslash}\vec{p}\cdot\vec{x}}\operatorname{and}\\
		u^{(2)} e^{\frac{\iun }{\hslash}\vec{p}\cdot\vec{x}}
	\end{cases}
\end{equation}
where momentum is given by $\vec{p}$ and energy by
$+\sqrt{m^2c^4+c^2p^2}$; and positron states by
\begin{equation}
	\begin{cases}
		u^{(3)} e^{\frac{\iun }{\hslash}\vec{p}\cdot\vec{x}}\operatorname{and}\\
		u^{(4)} e^{\frac{\iun }{\hslash}\vec{p}\cdot\vec{x}}
	\end{cases}
\end{equation}
with momentum equal to $-\vec{p}$ (and energy still as
$+\sqrt{m^2c^4+c^2p^2}$).
\end{minipage}

Given $\psi = \,u e^{\frac{\iun }{\hslash}\vec{p} \cdot \vec{x}}$
(with $u$ a four-component spinor), it is important to have two
operators $\mathcal{P}$ and $\mathcal{N}$ as \textit{projection
operators} such that $\mathcal{P}\psi$ contains only electron wave
functions and $\mathcal{N}\psi$ contains only negative energy wave
functions (positron states). As such, $\mathcal{P}$ and
$\mathcal{N}$ are spinor operators defined by:
\begin{equation}
	\begin{cases}
		\mathcal{P}u^{(1)} = u^{(1)}, \\
		\mathcal{P}u^{(2)} = u^{(2)}, \\
		\mathcal{P}u^{(3)} = 0, \\
		\mathcal{P}u^{(4)} = 0
	\end{cases}
\end{equation}
and
\begin{equation}
	\begin{cases}
		\mathcal{N}u^{(1)} = 0, \\
		\mathcal{N}u^{(2)} = 0, \\
		\mathcal{N}u^{(3)} = u^{(3)}, \\
		\mathcal{N}u^{(4)} = u^{(4)}.
	\end{cases}
\end{equation}
\noindent
These properties uniquely define $\mathcal{P}$ and $\mathcal{N}$.

\includepdf[pagecommand={}, height=\textheight, keepaspectratio, pages={8}]{Fermi.pdf}

\noindent
Observe that $Hu^{(1)} = Ru^{(1)}; Hu^{(2)} = Ru^{(2)};
Hu^{(3)} = -Ru^{(3)};$ and $Hu^{(4)} = -Ru^{(4)}$, with $R
= +\sqrt{m^2c^4 + c^2p^2}$ (here, $\vec{p}$ is a $c$-vector) and $H$
from (21). As such
\footnote{%
with the correction described in our section 4}
,
\begin{equation}
	\mathcal{P} = \frac{1}{2} + \frac{1}{2R}H; \;\;\;\;
	\mathcal{N} = \frac{1}{2} - \frac{1}{2R}H.
\end{equation}

\subsection*{Angular Momentum}

From (21), we obtain
\begin{equation}
	\left[
		H, xp_y - yp_x
	\right]
	=
	\frac{\hslash c}{\iun}
	\left(
		\alpha_1 p_y - \alpha_2 p_x
	\right)
	\neq 0.
\end{equation}

\noindent
Therefore, $xp_y - yp_x$ is not a time constant for a free Dirac
electron. However,
\begin{equation}
	xp_y - yp_x + \frac{1}{2} \frac{\hslash}{\iun} \alpha_1 \alpha_2 = \hslash J_z
\end{equation}
commutes with $H$. Interpret $\hslash J_z$ as the $z$
component of angular momentum:
\begin{equation}
	\hslash \vec{J}
	=
	\vec{x} \times \vec{p} +
	\frac{\hslash}{2{\iun}}
	\begin{cases}
		\alpha_2 \alpha_3 \\
		\alpha_3 \alpha_1 \\
		\alpha_1 \alpha2
	\end{cases}
	=
	\vec{x} \times \vec{p} +
	\frac{\hslash}{2}
	\vec{\sigma}'
\end{equation}
with $\vec{x} \times \vec{p}$ as the orbital part and
$\frac{\hslash}{2} \vec{\sigma}'$ as the spin part and
\begin{equation}
	\sigma_x' = \frac{1}{\iun} \alpha_2 \alpha_3 =
	\begin{vmatrix}
		0 & 1 & 0 & 0 \\
		1 & 0 & 0 & 0 \\
		0 & 0 & 0 & 1 \\
		0 & 0 & 1 & 0 \\
	\end{vmatrix};
	\;\;
	\sigma_y' = \frac{1}{\iun} \alpha_3 \alpha_1 =
	\begin{vmatrix}
		0 & -\iun & 0 & 0 \\
		\iun & 0 & 0 & 0 \\
		0 & 0 & 0 & -\iun \\
		0 & 0 & \iun & 0 \\
	\end{vmatrix};
	\;\;
	\sigma_z' = \frac{1}{\iun} \alpha_1 \alpha_2 =
	\begin{vmatrix}
		1 & 0 & 0 & 0 \\
		0 & -1 & 0 & 0 \\
		0 & 0 & 1 & 0 \\
		0 & 0 & 0 & -1 \\
	\end{vmatrix}.
\end{equation}

\noindent
Observe the analogy with the Pauli operators $\vec{\sigma}$ and
$\vec{\sigma}'$.

\vfill
\eject
\newpage

\setcounter{section}{9}

\section{Appendix C: Paper by Michel and Wightman\/}

This appendix contains, for the reader's convenience, the original text by Michel and Wightman of the abstract for their talk on a meeting
of the American Physical Society \cite{MicheWightman55} (Figure~3). Their result is cited in the \textquotedblleft Bible of Theoretical Physics\textquotedblright,
namely, in the L.~D.~Landau and E.~M.~Lifshitz Course of Theoretical Physics \cite{BerLifPit}.

\begin{figure}[h!]

                \includegraphics[width=0.77\linewidth]{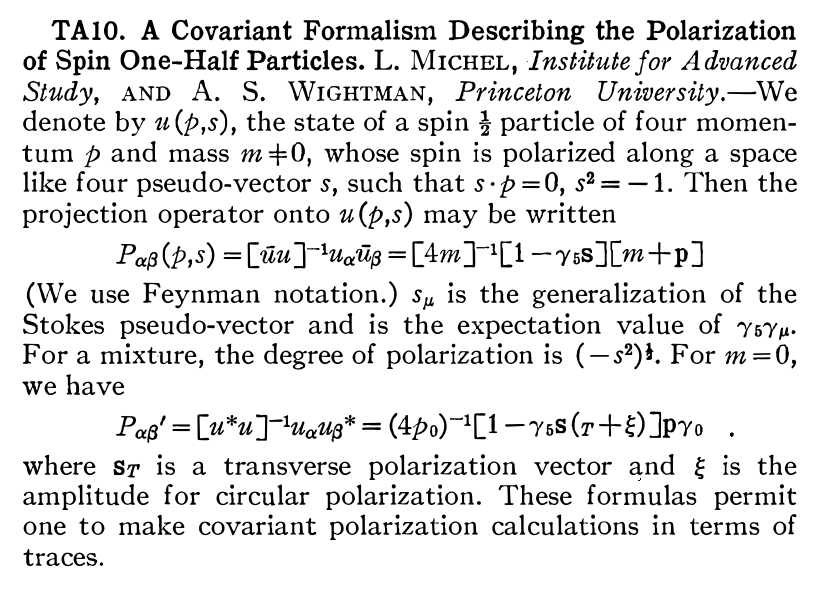}

                \caption{Reference \cite{MicheWightman55} }

\end{figure}

Our calculations in section~6 revealed the following connections with the
nonrelativistic polarization density matrices (\ref{s8aa})--(\ref{s8ab}):%
\begin{eqnarray}
&&\frac{u_{\alpha }^{\left( \pm 1/2\right) }\left( p\right) \overline{u}%
_{\beta }^{\left( \pm 1/2\right) }\left( p\right) }{\overline{u}^{\left( \pm
1/2\right) }\left( p\right) u^{\left( \pm 1/2\right) }\left( p\right) }=%
\frac{1}{2}\left(
\begin{array}{cc}
\mathbf{1}\pm \boldsymbol{\sigma \cdot n} & \mp \eta \left( \mathbf{1}\pm
\boldsymbol{\sigma \cdot n}\right)  \\
\pm \eta \left( \mathbf{1}\pm \boldsymbol{\sigma \cdot n}\right)  & -\eta
^{2}\left( \mathbf{1}\pm \boldsymbol{\sigma \cdot n}\right)
\end{array}%
\right)   \label{ac1} \\
&&\quad \quad =\left(
\begin{array}{cc}
\mathbf{1} & \mp \eta \mathbf{1} \\
\pm \eta \mathbf{1} & -\eta ^{2}\mathbf{1}%
\end{array}%
\right) \left(
\begin{array}{cc}
\rho ^{\left( \pm 1/2\right) }\left( \boldsymbol{\boldsymbol{n}}\right)  &
\mathbf{0} \\
\mathbf{0} & \rho ^{\left( \pm 1/2\right) }\left( \boldsymbol{\boldsymbol{n}}%
\right)
\end{array}%
\right) ,  \notag
\end{eqnarray}%
when $E>0$ and%
\begin{eqnarray}
&&\frac{v_{\alpha }^{\left( \mp 1/2\right) }\left( p\right) \overline{v}%
_{\beta }^{\left( \mp 1/2\right) }\left( p\right) }{\overline{v}^{\left( \mp
1/2\right) }\left( p\right) v^{\left( \mp 1/2\right) }\left( p\right) }=%
\frac{1}{2}\left(
\begin{array}{cc}
\eta ^{2}\left( \mathbf{1}\pm \boldsymbol{\sigma \cdot n}\right)  & \mp \eta
\left( \mathbf{1}\pm \boldsymbol{\sigma \cdot n}\right)  \\
\pm \eta \left( \mathbf{1}\pm \boldsymbol{\sigma \cdot n}\right)  & -\left(
\mathbf{1}\pm \boldsymbol{\sigma \cdot n}\right)
\end{array}%
\right)   \label{ac2} \\
&&\quad \quad =\left(
\begin{array}{cc}
\eta ^{2}\mathbf{1} & \mp \eta \mathbf{1} \\
\pm \eta \mathbf{1} & -\mathbf{1}%
\end{array}%
\right) \left(
\begin{array}{cc}
\rho ^{\left( \pm 1/2\right) }\left( \boldsymbol{\boldsymbol{n}}\right)  &
\mathbf{0} \\
\mathbf{0} & \rho ^{\left( \pm 1/2\right) }\left( \boldsymbol{\boldsymbol{n}}%
\right)
\end{array}%
\right) ,  \notag
\end{eqnarray}%
if $E<0.$ (In the nonrelativistic limit $\eta \to 0\/.$)

\end{document}